\documentclass[prl,aps,twocolumn,superscriptaddress,tightenlines,longbibliography]{revtex4-2}
\usepackage{graphicx}
\usepackage{bm}
\usepackage{braket}
\usepackage{pifont}
\usepackage{amsmath}
\usepackage{amsfonts}
\usepackage{amssymb}    
\usepackage{color,xcolor}
\usepackage[caption=false]{subfig}
\usepackage{tensor}
\usepackage{natbib}
\usepackage{dsfont}

 \newcommand{\beginsupplement}{%
        \setcounter{table}{0}
        \renewcommand{\thetable}{S\arabic{table}}%
        \setcounter{figure}{0}
        \renewcommand{\thefigure}{S\arabic{figure}}%
        \setcounter{equation}{0}
        \renewcommand{\theequation}{S\arabic{equation}}%
     }
\newcommand{\phantomsubfloat}[1]{
    {
        \captionsetup[subfigure]{labelformat=empty}
        \subfloat[][]{#1}
    }%
}
\newcommand {\mrm}[1] {\mathrm{#1}}

\newcommand{\JzA}{J^\text{z}_1}
\newcommand{\JzB}{J^\text{z}_2}
\newcommand{\JzC}{J^\text{z}_3}

\newcommand{\JMinusA}{J^-_1}
\newcommand{\JMinusB}{J^-_2}

\newcommand{\JMinusAB}{\JMinusA + \JMinusB}

\newcommand{\LR}[1]{\left< #1 \right>}
\begin{document}
\title{Reservoir-assisted energy migration through multiple spin-domains}
\author{Josephine Dias} 
\email{jdias@nii.ac.jp}
\thanks{These authors contributed equally}
\affiliation{National Institute of Informatics, 2-1-2 Hitotsubashi, Chiyoda, Tokyo 101-0003, Japan.}
\author{Christopher W. Wächtler} 
\email{cwaechtler@pks.mpg.de}
\thanks{These authors contributed equally}
\affiliation{NTT Basic Research Laboratories $\&$ NTT Research Center for Theoretical Quantum Physics, NTT Corporation,  3-1 Morinosato-Wakamiya,  Atsugi,  Kanagawa 243-0198,  Japan.}
\affiliation{Institut für Theoretische Physik, Technische Universität Berlin, Hardenbergstr. 36, 10623 Berlin, Germany}
\affiliation{Max Planck Institut für Physik komplexer Systeme, Nöthnitzer Str. 38, 01187 Dresden, Germany}
\author{Victor M. Bastidas}
\affiliation{NTT Basic Research Laboratories $\&$ NTT Research Center for Theoretical Quantum Physics, NTT Corporation,  3-1 Morinosato-Wakamiya,  Atsugi,  Kanagawa 243-0198,  Japan.}
\affiliation{National Institute of Informatics, 2-1-2 Hitotsubashi, Chiyoda, Tokyo 101-0003, Japan.}
\author{Kae Nemoto}
\affiliation{National Institute of Informatics, 2-1-2 Hitotsubashi, Chiyoda, Tokyo 101-0003, Japan.}
\author{William J. Munro}
\affiliation{NTT Basic Research Laboratories $\&$ NTT Research Center for Theoretical Quantum Physics, NTT Corporation,  3-1 Morinosato-Wakamiya,  Atsugi,  Kanagawa 243-0198,  Japan.}
\affiliation{National Institute of Informatics, 2-1-2 Hitotsubashi, Chiyoda, Tokyo 101-0003, Japan.}
\date{\today}
\begin{abstract}
The transfer of energy through a network of nodes is fundamental to both how nature and current technology operates. Traditionally we think of the nodes in a network being coupled to channels that connect them and then energy is passed from node to channel to node until it reaches its targeted site. Here we introduce an alternate approach to this where our channels are replaced by collective environments (or actually reservoirs) which interact with pairs of nodes. We show how energy initially located at a specific node can arrive at a target node - even though that environment may be at zero temperate. Further we show that such a migration occurs on much faster time scales than the damping rate associated with a single spin coupled to the reservoir. Our approach shows the power of being able to tailor both the system \& environment and the symmetries associated with them to provide new directions for future quantum technologies. 
\end{abstract}
\maketitle

\textit{Introduction--}
Nature has developed many methods for the transport of energy on length scales ranging from the atomic to cosmological \cite{krinner2017Transport,brown2019bad,Grohs2016,szydlowski2006cosmological}. Photosynthesis is one extremely well known example where pigment cofactors absorb the light and transfer it to antennae pigments where it is converted to chemical energy \cite{collini2010coherently,cao2020quantum,lambert2013quantum,kashida2018quantitative, franck1938migration, brixner2005two, christensson2012origin, ritschel2011efficient, briggs2011equivalence}. Such energy transport is not restricted to natural processes but is central to how our modern society and current technologies operate. We are always looking at new approaches to achieve this, but one needs to keep the possible applications in mind and the properties they require. In general, both classical and quantum systems are affected by the environment~\cite{BreuerPetruccioneBook2002} . The natural question here is: does noise help or hinder this transport process?  Actually (and counter intuitively) it was found that energy transport can be enhanced by adding environmental noise  \cite{rebentrost2009environment, plenio2008dephasing, gaab2004effects,biggerstaff2016enhancing,uchiyama2018environmental,Zhang2017}. Further, quantum mechanics provides unique opportunities in how energy transport could be enhanced using the principles of superpositions and entanglement ~\cite{einstein1935can,peres2006quantum,Matsuzaki2020}, and establishes tight bounds on how fast such energy transport processes can be~\cite{Lieb1972, deffner2017quantum,Tran2019}.  

The recent developments in quantum technology have given us excellent design options to tailor both our system and environmental properties to the tasks we want to achieve~\cite{verstraete2009quantum,barreiro2011open,Liu2020,seetharam2021}. It has been shown that a hybrid quantum systems composed of an ensemble of negatively charged nitrogen–vacancy (NV$^-$) centers in diamond coupled to a resonator~\cite{angerer2018superradiant} exhibits superradiant decay ~\cite{dicke1954coherence,gross1982superradiance} - a collective effect where radiation is amplified by the coherence of multiple emitters. In fact they showed collective decay twelve orders of magnitude faster than the decay of a single NV$^-$ center  \cite{angerer2018superradiant}. Interestingly the reverse process `superabsorption’  also exists - when radiation is absorbed much faster into the ensemble ~\cite{yang2021realization}, which has been experimentally realized by implementing a time-reversal process of superradiance~\cite{yang2021realization}. This was an extremely interesting observation from an energy transport point of view - combining the two phenomena would allow extremely fast energy transfer.

\begin{figure*}
\centering
 \includegraphics[width=0.8\linewidth]{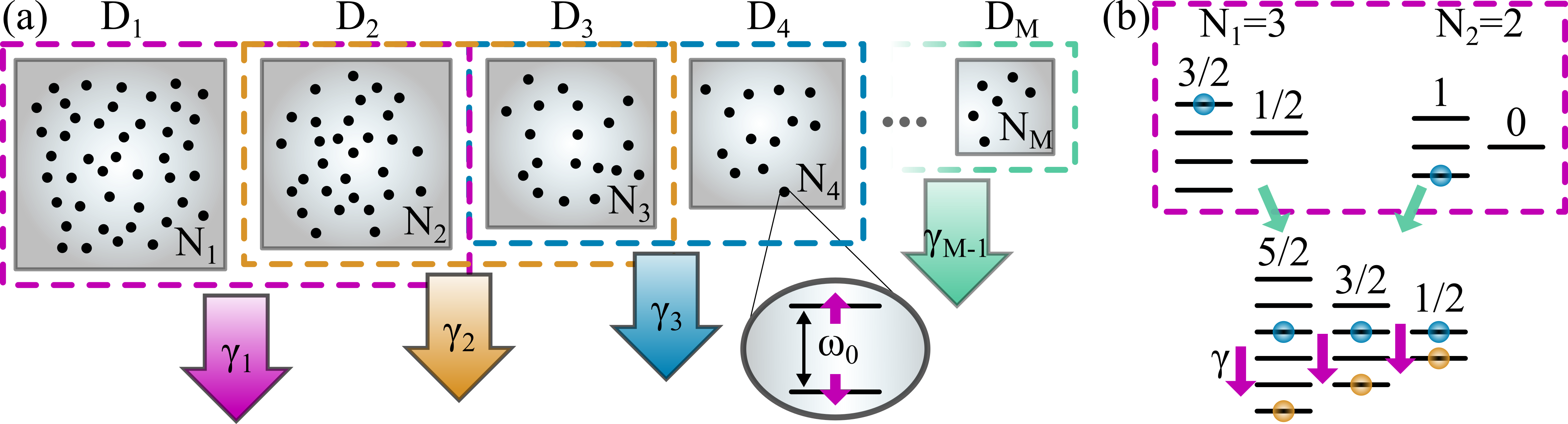}
\caption{a) Illustration of  a dissipatively coupled chain of spin domains. Each domain $D_i$ contains $N_i$ identical spin-1/2 particles.  All spin have an  energy $\hbar\omega_0$ with associated frequency frequency \(\omega_0 / 2 \pi\). Two neighboring domains $D_i$ and $D_{i+1}$ are interacting via a common (zero temperature) reservoir mediated by the dissipation rate $\gamma_i$. b) Schematic representation of the {\it effective} excitation transfer from the first domain (with $N_1=3$ spins and total angular momentum $j_1=3/2$ and $1/2$) to the second domain (with $N_2=2$ spins and total angular momentum $j_2=1$ and $0$): Initially $D_1$ is fully excited while $D_2$ is in the ground state.  This corresponds to the (partial) excitation of states with different total angular momentum (bottom row). Due to  the  collective decay, which preserves the  total angular momentum, the system relaxes down the ladders and reaches a steady state where $D_1$ and $D_2$ are locally not in their respective ground  states.  This results in excitations arising in the second domain - even though it was initially in its ground state.}
\label{fig:1}
\end{figure*}

One can however go further than this when one can engineer the environment ~\cite{verstraete2009quantum,Yanay2018,Keck2018,Damanet2019,Yanay2020}. Generally one would consider each ensemble coupling to its own  environment.  Hama et. al ~\cite{hama2018relaxation,hama2018negative} recently considered collective coupling of two ensembles to a reservoir and noted an unusual observation. They investigated what occurs if the first much larger ensemble was initially fully excited while the second is in its ground state and found  the first ensemble `superradiantly’ decays while the second ensemble undergoes `superabsorption’ (and can become fully populated). However the process is not that simple due to the nature of the coupling to the environment ~\cite{hama2018relaxation,fauzi2021}, that can induce coherent coupling~\cite{Astner2017,Norcia2018}.  Instead the key to explain the observed behavior lies in the symmetries of the system,  which can be seen from this very simple example. Consider two spins A and B initially in a state $|\psi\rangle=|1\rangle_A |0\rangle_B=|1\rangle |0\rangle$ which can also be expressed as 
\begin{eqnarray}
|\psi\rangle&= & \frac{1}{2} \left[ |1\rangle |0\rangle + |0\rangle |1\rangle\right] +  \frac{1}{2} \left[ |1\rangle |0\rangle - |0\rangle |1\rangle \right].
\end{eqnarray}
Under collective decay to a zero temperature bath, the first Bell state (a triplet state) decays to $|0\rangle |0\rangle $ while the second term (a dark state) remains unchanged. This means the mean population of spin B has increased from zero to $\bar n_B = 1/4$ via that collective coupling to the  reservoir. We must emphasize that there is no direct coupling between the spins meaning we are not seeing simple energy transfer.  Further those spins only collectively couple to a zero temperature reservoir meaning energy is not being given to the second spin from it. Instead this is a quantum process associated with the collective decay breaking a symmetry in the system and the symmetries of the initial state. The triplet (or symmetric) part of the initial state can decay but the dark (antisymmetric) part can not.  While this behavior can be seen in two spins \cite{minganti2021dissipative},  similar behavior can be seen with two ensembles collectively coupled to the environment \cite{hama2018relaxation,hama2018negative}.  We would like to highlight that this process is different from energy transfer in the traditional sense. Instead we call this energy migration to distinguish it. 

In this paper, our primary focus is to explore the fast migration of energy through a series of nodes - not connected by channels but collective environments instead. We utilize the well-known phenomenon of superradiance and superabsorption \cite{gross1982superradiance} to facilitate such energy migration between the multiple nodes. We will determine whether such a technique can be used to migrate energy around small networks where each node is an ensemble of spins that are collectively coupled to an environment. 

\textit{Our Model--}
Let us begin with a simple mathematical model of our system which extends a double domain system ~\cite{hama2018relaxation, hama2018negative} to the multiple spin domain regime.  Our system depicted in  Fig.~\ref{fig:1}(a) consists of $M$ different non-interacting spin domains $D_i$, each containing $N_i$ identical spin-1/2 particles (with frequency \(\omega_0 / 2 \pi\)). Pairwise these domains are collectively coupled to a zero temperature reservoir.  These reservoirs are modeled as a collection of bosonic modes with frequencies $\omega_{k_i} / 2 \pi$ and bosonic creation (annihilation) operators $a^\dagger_{k_i}$ ($a_{k_i}$). Importantly, our system is symmetric under exchange of any two spins within each domain but not within the overall system. Therefore, it is useful to define collective spin operators for the $i^{th}$ domain $J_i^\alpha=\sum_{n_i=1}^{N_i} S_{n_i}^\alpha$  with $\alpha=\text{x}, \text{y},\text{z}$ and where $S_{n_i}^\alpha$ are the $n_i$th spin operators.  Further the $i^{th}$ domain raising and lower operators are given by  $J_i^{\pm}=J_i^\text{x}\pm \text{i}  J_i^\text{y}$. The Hamiltonian of the total system with $M$ ensembles and $M-1$ reservoirs is
\begin{eqnarray}
H&=& \hbar \omega_0  \sum\limits_{i}^M J_i^z+ \sum\limits_{i}^{M-1} \sum\limits_{k_i} \hbar \omega_{k_i} a_{k_i}^\dagger a_{k_i} \\
&+&\sum\limits_{i}^{M-1} \sum\limits_{k_i} \left[t_{k_i}\left(J_i^+ + J_{i+1}^+\right)a_{k_i} + t_{k_i}^\ast a_{k_i}^\dagger\left(J_i^- + J_{i+1}^-\right)\right], \nonumber
\label{eq:ham}
\end{eqnarray}
where the first and second term represents the Hamiltonian of the spin ensembles and the bosonic reservoirs, respectively.  The third term is the interaction of the spin ensemble $i$ and $i+1$ with their common reservoir (labeled as $i$) where $t_{k_i}$, $t_{k_i}^\ast$ represent emission (absorption) amplitudes that fix the spectral density of the reservoirs $\Gamma_i(\omega) = 2\pi \sum_{k_i} |t_{k_i}|^2\delta(\omega-\omega_{k_i})$. Within the standard weak-coupling approach (Born-Markov approximation) and assuming zero-temperature reservoirs, the Lindblad master equation of the system can be written as \cite{BreuerPetruccioneBook2002, carmichael2013statistical}:
\begin{align}
\begin{split}
\label{eq:master}
\dot{\rho}_s= -i \omega_0 &  \sum\limits_i^M\left[ J_i^z, \rho_s \right] + \sum\limits_i^{M-1}\frac{\gamma_{i}}{2}\mathcal{D}\left[J_i^-+J_{i+1}^-\right]\rho_s
\ ,
\end{split}
\end{align}
where the Lindblad term is $\mathcal{D}\left[O\right]\rho = 2 O \rho O^\dagger-O^\dagger O \rho- \rho O^\dagger O$ for any operator $O$. The dissipative coupling between the different spin ensembles is mediated via the rates $\gamma_i = \Gamma_i(\omega_0) = \alpha_i \omega_0$ in the wide band limit, where $\alpha_i $ is constant. Previous works have used dissipative coupling to induce frustration~\cite{Li2021} and quantum  synchronization of oscillators~\cite{Lee2014} and atomic ensembles~\cite{Xu2014}.

\begin{figure*}
\centering
\captionsetup[subfigure]{labelformat=empty}
\subfloat[]{\includegraphics[width=0.375\linewidth]{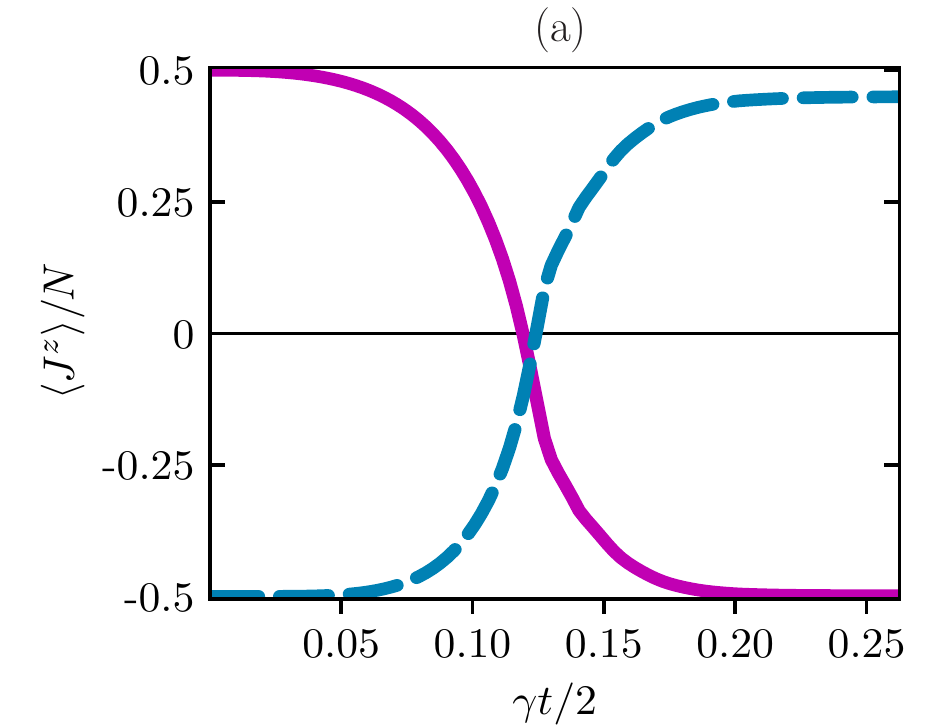} \label{fig:d2}}\subfloat[]{\includegraphics[width=0.3125\linewidth]{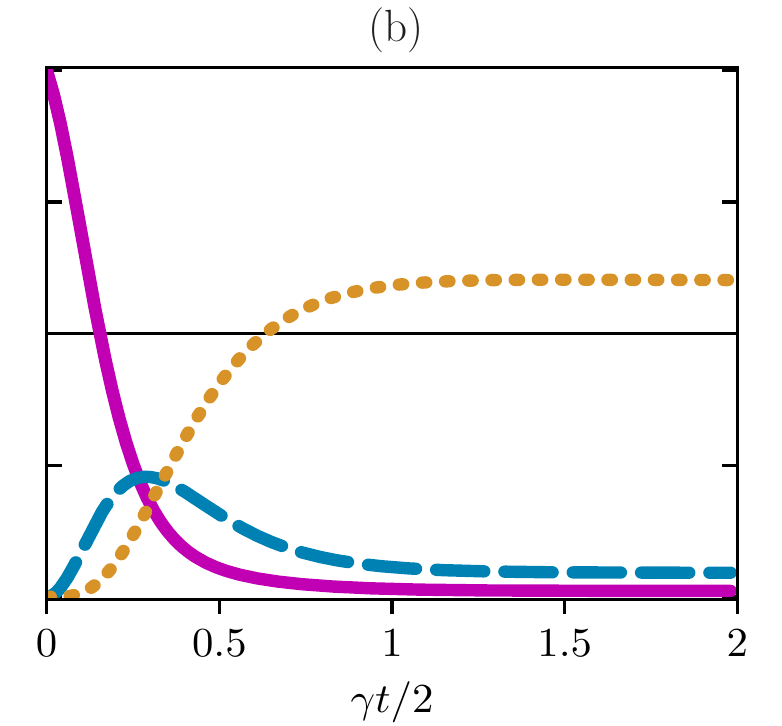}\label{fig:d3}}\subfloat[]{\includegraphics[width=0.3125\linewidth]{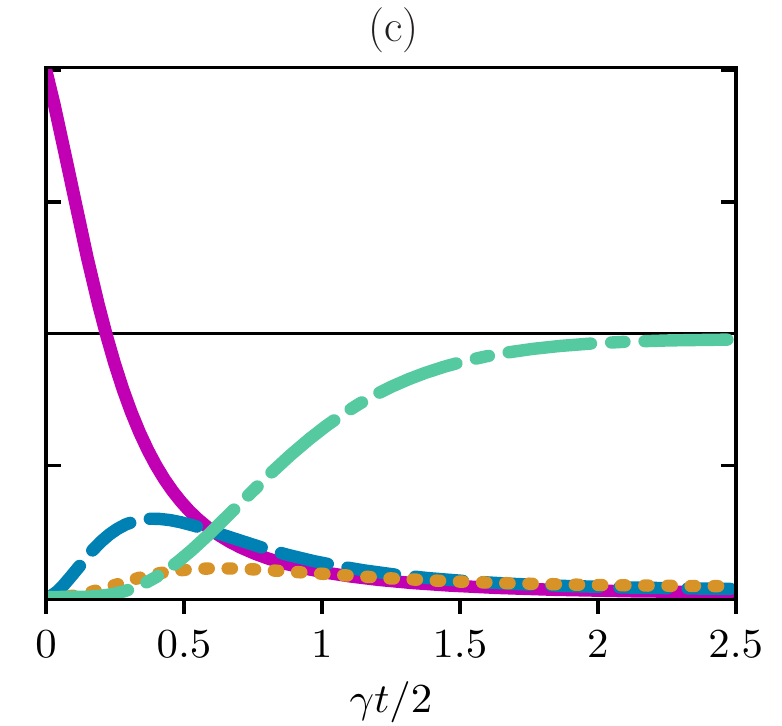}\label{fig:d4}}
\vspace{-2\baselineskip}
\caption{Normalized collective spin relaxation $\langle J^z_i\rangle/N_i$ with dynamics governed by the master equation~(\ref{eq:master}). \protect\subref{fig:d2} Two spin domains with $N_1=40$ (solid magenta), $N_2=2$ (dashed blue), \protect\subref{fig:d3} three domains with  $N_1=12$ (solid magenta), $N_2=6$ (dashed blue), $N_3=2$ (dotted gold),  \protect\subref{fig:d4} four domains with  $N_1=8$ (solid magenta), $N_2=6$ (dashed blue), $N_3=5$ (dotted gold) and $N_4=1$ (dot-dashed mint green). }
\label{fig:2}
\end{figure*}

\textit{System dynamics--}
As we are mostly interested in excitation migration through the different spin domains, our results will focus on the situations where the first domain is initialized with all spins in their excited states while all spins in subsequent domains begin in the ground state.  Our initial state can be expressed as
\begin{equation}
\ket{\mrm{is}}=\ket{\uparrow...\uparrow}_1 \otimes \ket{\downarrow...\downarrow}_2 \otimes \dots \otimes \ket{\downarrow...\downarrow}_{M}
\ .
\label{eq:initialState}
\end{equation}
As observed in \cite{hama2018negative} for the two domain case, this initial state~\eqref{eq:initialState} is not symmetric under exchange of the ensembles 1 and 2. It is worth noting that the dissipative terms in the master equation~(\ref{eq:master}) induce correlations between neighboring domains $i$ and $i+1$, because it describes their collective decay. The initial state can be decomposed as a superposition of symmetric and antisymmetric states. As a consequence of the decay of the symmetric subspace components, the average number of excitations stays finite at the steady state despite the presence of the zero temperature reservoir [see Fig.~\ref{fig:1}(b)]. Furthermore, the system may relax into a steady state, where the second domain population of spins in the excited state is greater than $50\%$. However, an unbalanced configuration of domain sizes -- specifically $N_1\gg N_2$ -- is necessary for this situation to occur. For this reason, we restrict ourselves to the unbalanced configurations where $N_1>N_2>\dots >N_M$ and explore the dynamics of excitation migration from the first to the last domain. 

We are now in the position to explore the dynamics of the dissipatively coupled spin ensembles. In Figs.~\ref{fig:d2}--\ref{fig:d4} we show the collective spin relaxation of a system with two, three and four domains with initial state $\ket{\mrm{is}}$ given by \eqref{eq:initialState}. For the two domain system shown in  Fig.~\ref{fig:d2} we set $N_1=40$ (magenta) and $N_2=2$ (blue),  while for the three domain system [Fig.~\ref{fig:d3}] we have $N_1=12$ (magenta), $N_2=6$ (blue) and $N_3=2$ (gold).  Similarly for the four domain system shown in Fig.~\ref{fig:d4} we have $N_1=8$ (magenta), $N_2=6$ (blue), $N_3=5$ (gold) and $N_4=1$ (mint green). Since  we solve Eq.~\eqref{eq:master} numerically, we are restricted to rather small ensemble sizes especially as the number of spin domains increases. Nevertheless, small systems provide valuable insights into the general dynamics and allows to draw conclusions.

It can be seen in Fig.~\ref{fig:2} that decay of the first domain first leads to excitation of the second domain as a result of the reservoir-mediated interaction between the two domains. In the case of only two domains [Fig.~\ref{fig:d2}], the dynamics comes to a halt and  the  system reaches a steady state. Due to the large imbalance $N_1>N_2$, the first domain (magenta) is (almost) completely deexcited and the second domain (blue) gets close to the fully excited state \cite{hama2018negative}. For more than two domains, the second domain (at a slower rate) also decays due to the additional dissipation channel and the ensemble excitation is transferred to the third domain [Fig.~\ref{fig:d3}]. This process will continue until the last (smallest) domain absorbs the excitation and the system finally reaches its steady state solution where the smallest ensemble is excited. For the system sizes considered here, the intermediate domain population of spins in the excited state stays below $50\%$ and the average number of excitations in the last domain is considerably less than the ensemble size. However, the results of the two domain system depicted in Fig.~\ref{fig:d2} indicate that larger ensemble sizes may allow to dissipatively migrate the fully excited initial state from the first domain along the chain to the last domain. 
\begin{figure}
\centering 
\includegraphics[width=0.98\linewidth]{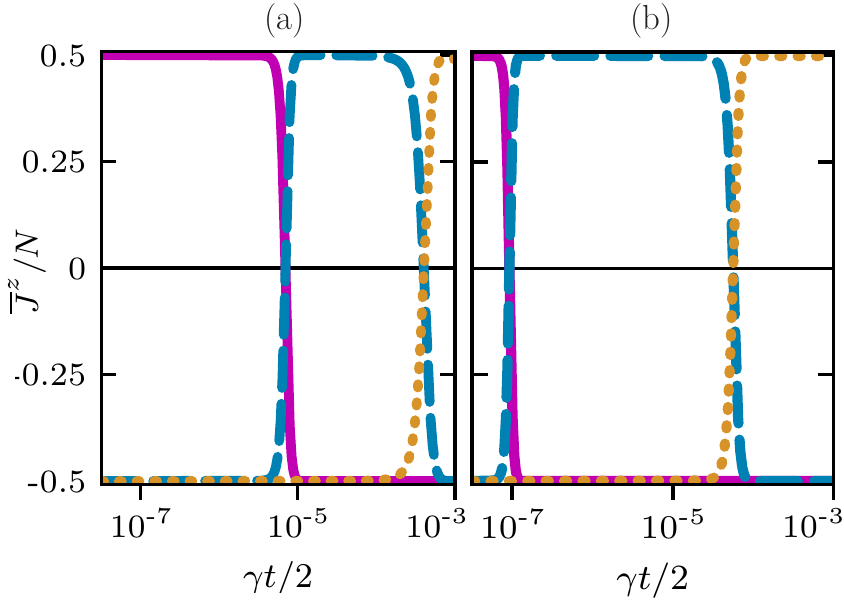}
\phantomsubfloat{\label{fig:mf1}}\phantomsubfloat{\label{fig:mf2}}
\vspace{-2\baselineskip}
\caption{Normalized collective spin relaxation $\langle J^z_i\rangle/N_i$ for three domains using mean-field dynamics. \protect\subref{fig:mf1} $N_1=10^6$ (solid magenta), $N_2=10^4$ (dashed blue) and $N_3=10^2$ (dotted gold). \protect\subref{fig:mf2}  $N_1=10^8$ (solid magenta), $N_2=10^5$ (dashed blue), $N_3=10^2$ (dotted gold).}\label{fig:3}
\end{figure}
\begin{figure}
\centering \includegraphics[width=0.98\columnwidth]{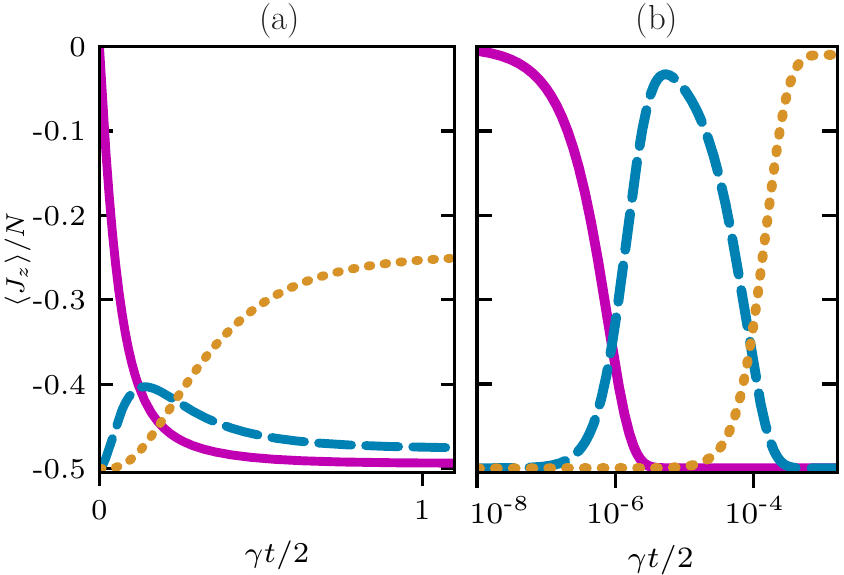}
\phantomsubfloat{\label{fig:alt1}}\phantomsubfloat{\label{fig:alt2}}
\vspace{-2\baselineskip}
\caption{Collective spin relaxation $\langle J^z_i\rangle/N_i$ for an initial state of the first domain with $\braket{J^z_1}\left(t=0\right)=0$ showing that the relative population of the later domains never exceeds  $\braket{J^z_{2,3}}=0$ . \protect\subref{fig:alt1} The master equation approach with $N_1=12$ (solid magenta), $N_2=6$ (dashed blue),  $N_3=2$ (dotted gold). \protect\subref{fig:alt2} The mean field approach with $N_1=10^6$ (solid magenta),  $N_2=10^4$ (dashed blue),  $N_3=10^2$ (dotted gold).}\label{fig:alt}\end{figure}

The dynamics we are able to access from numerically solving Eq.~\eqref{eq:master} are intriguing, however due to the scaling of the Hilbert space with system size, we are limited in the number of spins. In order to investigate larger domain sizes, we perform a mean-field (MF) approximation by factorizing different moments (see Supplemental Material). In the following, we consider the dynamics of three dissipatively coupled spin domains which is described by a closed set of $13$ coupled differential equations for the expectation values of the collective spin operators, which we denote by an overbar to emphasize their MF character. 

In Fig.~\ref{fig:3}, we show the collective spin relaxation according to the MF description for two different ensemble sizes of the first and second domain. In Fig.~\ref{fig:mf1} we set $N_1=10^6$ (magenta), $N_2=10^4$ (blue) and $N_3=10^2$ (gold), and in Fig.~\ref{fig:mf2}  $N_1=10^8$ (magenta), $N_2=10^5$ (blue)  and $N_3=10^2$ (gold). Unlike the smaller systems sizes shown in Fig.~\ref{fig:2}, each domain differs substantially in the number of spins from the previous domain. Consequently, we are able to witness (almost) full excitation of the second and third domains,  with $D_3$ relaxing to a steady state of $\bar J_3^z/N_3\approx 0.49$. This is in contrast to the results of Fig.~\ref{fig:d3}, where $\left<J_2^z\right>/N_2$ and $\left<J_3^z\right>/N_3$ clearly do not reach the maximum of $\left<J_1^z\right>/N_1 \left(t=0\right)=1/2$. 

Additionally, the discrete migration of excitation from one spin-domain to the next can clearly be witnessed (notice the logarithmic time scale), in contrast to the collective spin relaxations of Fig.~\ref{fig:2} where all domains absorb and decay on the same time scale. This is a clear signature of the superradiant decay and thus superradiant excitation transfer, which scales as $1/N_i$ for large system sizes. From the dynamics shown in Fig.~\ref{fig:3}, we note that the superradiant decay time of $D_1$ as well as the superradiant absorption time of $D_2$ is mostly governed by the size of $D_1$, thus occurs at a time that is orders of magnitude before the superradiant decay of $D_2$ and absorption of $D_3$. Therefore,  efficient migration in this dissipatively coupled system occurs when the spin population of the initial excited domain ($N_1$) is sufficiently larger than the final domain (in this case $N_3$). 

So far, we have explored the superradiant migration of excitations when the first domain is initially fully excited and all subsequent domains are in their respective ground state. However, one may assume that a partially excited initial state is sufficient to fully excite the last domain as the number of spins within each domain decreases along the chain. In the following, we show that this is in fact not the case and, moreover, that the maximum relative excitation transferable from one domain to the other is bounded by the initial relative excitation. 

Let us start by numerically exploring the three domain system where small and large system sizes can be investigated. We first study the effect of initial conditions on the excitation transfer described by the master equation \eqref{eq:master} for domain sizes $N_1=12$, $N_2=6$ and $N_3=2$. As $N_2=N_1/2$ we choose as initial state the first domain to be only half excited ($\left<J_1^\text{z}\right>(t=0)=0$), however, the second and third domain to be in their respective ground state. In Fig.~\ref{fig:alt1} one sees that for this initial configuration the second (blue) and third domain (gold) are both less excited compared to the initial state $\ket{\mrm{is}}$ [cf. Fig.~\ref{fig:d3}], and, especially, the third domain is considerably below half excited. In contrast, in Fig.~\ref{fig:alt2},  the spin relaxation dynamics is shown for $N_1=10^6$, $N_2=10^4$ and $N_3=10^2$ [same system sizes as Fig.~\ref{fig:mf1}] with half of the spins in $D_1$ initialized in the excited state and half in the ground state. Here, we make use of the MF equations to solve the dynamics. Even though the number of spins in the second and third domains are significantly less than the number of initially excited spins in the first domain,  both $\bar J_2^z/N_2$ and $\bar J_3^z/N_3$ remain always below the value of $ \bar J_1^z/N_1 \left(t=0\right)$.  Interestingly,  this occurs for any value of $N_1$ and any proportion of excited spins in the initial state.  That is, we observe $\bar J_1^z/N_1 \left(t=0\right) \geq \max\left(\bar J_2^z/N_2\right)$ and  $\bar J_1^z/N_1 \left(t=0\right) \geq \max\left(\bar J_3^z/N_3\right)$. This already suggests that the initial population of spins in the excited state limits the transferable amount of excitations. 

The results we have observed so far for the three domain case, also holds for the case of only two domains. That is the the maximum relative excitation of the second domain is bounded by the initial relative excitation of the first domain. In fact, as we show in the Supplemental Material, for the two domain case with $N_1\gg N_2$ the steady state of the second domain is approximately given by 
\begin{equation}
\label{eq:MaximumExciation}
\frac{\bar \JzB}{N_2}(t\to\infty)\approx\frac{\bar\JzA}{N_1}(t=0)
\ .
\end{equation}
As we discussed in the previous section, for superradiant excitation transfer to  occur we need large differences in the number of spins within each domain. This results in a time scale separation of transfer between the first and second and transfer between the second and  third domain. Because of this time scale separation the second domain reaches its maximum relative excitation before transport to the third domain takes place. We thus conclude from Eq.~\eqref{eq:MaximumExciation} that the maximum relative excitation of the last domain is bounded by the initial state, i.e., ${\bar J_\text{M}^z}/{N_M}(t\to\infty)\approx{\bar\JzA}/{N_1}(t=0)$. This has implications for quantum thermodynamics and especially the charging of quantum batteries~\cite{quach2020using}.

\textit{Discussion--}
It is well established that movement or transfer of energy around physical system is a primitive operation with applications in many diverse fields. We are always looking for new ways to achieve this in faster and more efficient ways. In this article we have shown an energy migration approach in a small scale quantum network based on collective coupling to a reservoir. Energy is not flowing from node to node. Instead our initial state is not symmetric with respect to the collective coupling to the reservoirs and so different parts of the quantum wavefunction decay at different rates (or not at all). This results in populations arising in nodes which we initially unoccupied.  Combining this behavior with superradiant decay and absorption, we show the apparent flow of energy from node to node in the network. In exploring the dynamics of energy migration in the network, we were able to find the conditions which facilitate the fastest and most efficient energy transfer.  By tailoring the system and environment and symmetries associated with them, our approach can illustrate new directions for the future of quantum technologies.

\textit{Acknowledgments--} We thank A. Eisfeld and A. Sakurai for useful discussions.  This work was supported by the JSPS KAKENHI Grant No. 19H00662 and the MEXT Quantum Leap Flagship Program (MEXT QLEAP) Grant No.  JPMXS0118069605.  CWW acknowledges  financial support from the Deutsche Forschungsgemeinschaft through Project No. BR1528/8-2 and from the Max-Planck Gesellschaft via the MPI-PKS Next Step fellowship.

\clearpage
\onecolumngrid
\appendix
\beginsupplement
\section*{SUPPLEMENTAL MATERIAL}
\setcounter{section}{0}
\renewcommand{\thesection}{\Roman{section}}
\renewcommand{\thesubsection}{\thesection.\Roman{subsection}}
\section*{I. Mean-field approximation \label{app:meanfield}}
In this section we apply a MF approximation to the three domain system  to access the system dynamics for large domain sizes. The full dynamics is described by Eq.~(3) with $M=3$. In order to obtain a consistent closed set of equations, we make the following approximations:
\begin{equation}
\label{eq:AppMF}
\begin{aligned}
\left<\left(J^\text{z}_i\right)^2\right>&\approx\left<J^\text{z}_i\right>^2,\quad \left<\left(J^\text{z}_i\right)^2 J^\text{z}_j\right> \approx \left<J^\text{z}_i\right>\left<J^\text{z}_i J^\text{z}_j\right>,\quad \left<J^\text{z}_\text{1/2} A\right>\approx\left<J^\text{z}_\text{1/2}\right>\left<A\right>,  \\ \left<J^\text{z}_\text{2/3} B\right>&\approx\left<J^\text{z}_\text{2/3}\right>\left<B\right>,\quad  
\left<J^\text{z}_\text{1/3} C\right>\approx\left<J^\text{z}_\text{1/3}\right>\left<C\right>,
\end{aligned}
\end{equation}
where we defined the new operators 
\begin{equation}
A = J^+_1 J^-_2+J^-_1 J^+_2,\qquad B = J^+_2 J^-_3+J^-_2 J^+_3, \qquad C = J^+_1 J^-_3+J^-_1 J^+_3. 
\end{equation}  
Note that we do not factorize expectation values that involve operators of all three domains, e.g. $\left<J^\text{z}_1 B\right>$. 

Using the approximations \eqref{eq:AppMF}, we find in total $13$ coupled equations of motion:
\begin{equation}
\begin{aligned}
\frac{d}{dt}\left<\JzA\right> =& -\gamma\left[\bar N_1-\left<\JzA\right>^2+\left<\JzA\right>+\frac{1}{2}\left<A\right>\right], \\
\frac{d}{dt}\left<\JzB\right> =& -\gamma\left[\bar N_2-\left<\JzB\right>^2+\left<\JzB\right>+\frac{1}{2}\left<A\right>\right] -\gamma\left[\bar N_2-\left<\JzB\right>^2+\left<\JzB\right>+\frac{1}{2}\left<B\right>\right],\\
\frac{d}{dt}\left<\JzC\right> =& -\gamma\left[\bar N_3-\left<\JzC\right>^2+\left<\JzC\right>+\frac{1}{2}\left<B\right>\right], \nonumber
\end{aligned}
\end{equation}
\begin{equation}
\label{eq:MF3Systems1}
\begin{aligned}
\frac{d}{dt}\left<A\right> =&\gamma \left[\left<\JzA\right>\left(\left<A\right>  +2\bar N_2-2\LR{\JzA \JzB}\right) + \left<\JzB\right>\left(2\left<A\right> +2 \bar N_1 -2 \LR{\JzA \JzB}\right)+4 \LR{\JzA \JzB}+\LR{\JzB C} - \frac{3}{2}\LR A\right],\\
\frac{d}{dt}\left<B\right> =&\gamma \left[2\left<\JzB\right>\left(\left<B\right>  +2\bar N_3-2\LR{\JzB \JzC}\right) + \left<\JzC\right>\left(\left<B\right> +2 \bar N_2 -2 \LR{\JzB \JzC}\right)+4 \LR{\JzB \JzC}+\LR{\JzB C} - \frac{3}{2}\LR B\right],\\
\frac{d}{dt}\left<C\right> =& \gamma \left[\LR{\JzA}\LR{C} + \LR{\JzA B} - \frac{1}{2}\LR C\right]+ \gamma \left[\LR{\JzC}\LR{C} + \LR{\JzC A} - \frac{1}{2}\LR C\right],  
\end{aligned}
\end{equation}
\begin{equation}
\begin{aligned}
\frac{d}{dt}\LR{\JzA \JzB} =& -\frac{\gamma}{2} \left<A\right>\left(\left<\JzA\right> + \left<\JzB\right>-1\right) -\gamma \bar N_1 \LR \JzB - 2\gamma \bar N_2\LR \JzA+\gamma \LR{\JzA \JzB}\left(\LR{\JzA} + 2\LR{\JzB} - 3\right)-\frac{\gamma}{2}\LR{\JzA B},\\
\frac{d}{dt}\LR{\JzB \JzC} =& -\frac{\gamma}{2} \left<B\right>\left(\left<\JzB\right> + \left<\JzC\right>-1\right) -\gamma \bar N_3 \LR \JzB - 2\gamma \bar N_2\LR \JzC+\gamma \LR{\JzB \JzC}\left(\LR{\JzB} + 2\LR{\JzC} - 3\right)-\frac{\gamma}{2}\LR{\JzC A},\\
\frac{d}{dt}\LR{\JzA \JzC} =& - \gamma \left[\LR \JzC \bar N_1 - \LR{\JzA\JzC}\LR \JzA\right] -\gamma \left[\LR{\JzA\JzC} + \frac{1}{2}\LR{\JzC A}\right].\\
\frac{d}{dt}\LR{\JzA B} =& -\frac{\gamma}{2} \left[2\bar N_1 \LR B +\LR A \LR B\right] +\frac{\gamma}{2} \LR{\JzA B}\left[2\LR \JzA +4\LR \JzB +2\LR \JzC -5\right] +\gamma \LR{\JzB C}\left[\LR \JzA -1\right] \\
&+2\gamma\LR{\JzA\JzB\JzC}\left[2-\LR \JzB-\LR \JzC\right] +2\gamma \LR{\JzA\JzB} \bar N_3 + 2\gamma\LR{\JzA\JzC} \bar N_2,\\
\frac{d}{dt}\LR{\JzC A} =& -\frac{\gamma}{2}\left[2\bar N_3 \LR A +\LR A \LR B\right] +\frac{\gamma}{2} \LR{\JzC A}\left[2\LR \JzC +4\LR \JzB +2\LR \JzA -5\right] +\gamma \LR{\JzB C}\left[\LR \JzC -1\right] \\
&+2\gamma\LR{\JzA\JzB\JzC}\left[2-\LR \JzA-\LR \JzB\right]+2\gamma \LR{\JzB\JzC} \bar N_1+ 2\gamma\LR{\JzA\JzC} \bar N_2,\\
\frac{d}{dt}\LR{\JzB C} =& \gamma \left[\LR{\JzB C}\left(\LR \JzA +2\LR \JzB +\LR \JzC-3\right) + \left(\LR{\JzA B}+\LR{\JzC A}\right)\left(\LR \JzB -1\right) -\LR{C} \left(2\bar N_2+\frac{1}{2}\LR A+\frac{1}{2}\LR B\right)\right],\\
\frac{d}{dt}\LR{\JzA \JzB\JzC} =& -\gamma\left[ \bar N_1 \LR{\JzB\JzC} +2\bar N_2\LR{\JzA\JzC} + \bar N_3\LR{\JzA\JzB}\right]+\gamma \LR{\JzA \JzB\JzC}\left[\LR \JzA +2\LR \JzB +\LR \JzC -4\right] \\
&-\frac{\gamma}{2} \LR{\JzA B}\left[\LR \JzB+\LR \JzC -1\right] -\frac{\gamma}{2} \LR{\JzC A}\left[\LR \JzA+\LR \JzB -1\right] \nonumber
\end{aligned}
\end{equation}
where we have defined $\bar N_i\equiv \left({N_i}/{2}\right)\left({N_i}/{2}+1\right)$. The set of equations defined in Eq.~\eqref{eq:MF3Systems1} describes the dynamics of the three domain system in the limit of large numbers of spin within in each domain. 

\section*{II. Maximum relative excitation of the second domain}
\label{App:Bound}

In this section we examine the two domain case and investigate the effect of the initial relative excitation of domain $D_1$ on the maximum final excitation of domain $D_2$. Specifically, we are interested in the steady state limit $\left<\JzB\right>(t\to\infty)$. To gain some initial insight, let us start with small system sizes where the first domain consists of $N_1=4$ spins, initially all excited while the second domain is composed of a single spin in the ground state. The corresponding state can be written as a tensor product
\begin{equation}
\ket{\psi}(t=0)=\ket{j_1~m_1}\otimes\ket{j_2~m_2}=\ket{2~2}\otimes\ket{\frac{1}{2}~-\frac{1}{2}}.
\end{equation}
Since the operator $\JMinusAB$ is a \emph{total angular momentum} operator, it is useful to change to the basis of eigenstates of the total angular momentum. In this basis, the initial state can be expressed using the Clebsch-Gordan coefficients as
\begin{equation}
\ket{\psi}(t=0)=\frac{1}{\sqrt{5}}\ket{\frac{5}{2}~\frac{3}{2}} + \sqrt{\frac{4}{5}}\ket{\frac{3}{2}~\frac{3}{2}}.
\end{equation}
The evolution of this state is only dissipative according to the master equation. Hence, the final state will be 
\begin{equation}
\ket{\psi}(t\to\infty)=\frac{1}{\sqrt{5}}\ket{\frac{5}{2}~-\frac{5}{2}} + \sqrt{\frac{4}{5}}\ket{\frac{3}{2}~-\frac{3}{2}}.
\end{equation}
Now, in order to calculate $\left<\JzB\right>(t\to\infty)$, we need to change our basis back to the original one,
\begin{equation}
\ket{\psi}(t\to\infty)=\frac{1}{\sqrt{5}}\ket{2~-2}\otimes\ket{\frac{1}{2}~-\frac{1}{2}} + \sqrt{\frac{4}{5}}\left(\frac{1}{\sqrt{5}}\ket{2~-1}\otimes\ket{\frac{1}{2}~-\frac{1}{2}}-\sqrt{\frac{4}{5}}\ket{2~-2}\otimes\ket{\frac{1}{2}~\frac{1}{2}}\right)
\end{equation}
from which we establish $\left<\JzB\right>(t\to\infty)={7}/{50}$.

We can generalize the previous calculation to the case of $N_1$ spins up domain $D_1$ and one single spin in the ground in the second domain. In such a case \cite{hama2018negative2}
\begin{equation}
\left<\JzB\right>(t\to\infty) = \frac{N_1^2-2 N_1-1}{2\left(N_1+1\right)^2}. 
\end{equation}
which tends to $1/2$ as $N_1 \rightarrow \infty$. That is,
the second domain $D_2$ becomes fully exited under collective relaxation. Now what about the situation where $D_1$ is not initially fully excited ($\left<\JzA\right>(t=0)\neq N_1/2$). In this case, it is straightforward to show that
\begin{equation}
\left<\JzB\right>(t\to\infty) = \frac{\left<\JzA\right>(t=0) N_1- N_1-\frac{1}{2}}{\left(N_1+1\right)^2},
\end{equation}
which in the limit of large $N_1$ becomes
\begin{equation}
\lim\limits_{N_1 \to\infty}\frac{\left<\JzB\right>(t\to\infty)}{1} = \frac{\left<\JzA\right>}{N_1}(t=0).
\end{equation}

So far we have only examined the case of a single spin in the second domain. However, using the approximate MF equations we can explore this regime. In the case of only two domains, Eqs.~(\ref{eq:MF3Systems1}) reduce to 
\begin{equation}
\label{eq:MF2Systems}
\begin{aligned}
\frac{d}{dt}\left<\JzA\right> =& -\gamma\left[\bar N_1-\left<\JzA\right>^2+\left<\JzA\right>+\frac{1}{2}\left<A\right>\right], \\
\frac{d}{dt}\left<\JzB\right> =& -\gamma\left[\bar N_2-\left<\JzB\right>^2+\left<\JzB\right>+\frac{1}{2}\left<A\right>\right],\\
\frac{d}{dt}\left<A\right> =&\gamma \left[\left<\JzA\right>\left(\left<A\right>  +2\bar N_2-2\LR{\JzA \JzB}\right)- \LR A\right] +\gamma\left[ \left<\JzB\right>\left(\left<A\right> +2 \bar N_1 -2 \LR{\JzA \JzB}\right)+4 \LR{\JzA \JzB} \right],\\
\frac{d}{dt}\LR{\JzA \JzB} =& -\frac{1}{2}\frac{d}{dt}\LR A.
\end{aligned}
\end{equation}
As initial conditions we assume $\left<\JzB\right>(0) = -N_2/2$ and $\left<A\right>(0)=0$, i.e., no initial correlations between the first and second domain. 

At steady state ($t\to \infty$) the left hand side of the above equations is equal to zero. If we furthermore divide the second line of Eq.~(\ref{eq:MF2Systems}) by $N_1$ and assume $N_1\gg N_2$ (or more specifically $N_1\gg N_2^2$), we arrive at $\left<A\right>(t)/N_1 \approx 0$ because the absolute value of $\left<\JzB\right>(t)$ is bounded by the number of spins $N_2$. In this case, we can apply an additional mean field approximation of the form $\LR{\JzA \JzB} \approx \LR \JzA \LR \JzB$ such that from the last line of Eq.~(\ref{eq:MF2Systems}) it follows that $\LR A (t) = -2\LR \JzA (t)\LR \JzB (t) + 2\LR \JzA (0)\LR \JzB (0)$. Here, the second term ensures that the initial condition $\left<A\right>(0)=0$ is fulfilled. Remembering the initial condition $\left<\JzB\right>(0) = -N_2/2$, we have 
\begin{equation}
\frac{\left<\JzB\right>}{N_2}(t\to\infty)=\frac{\left<\JzA\right>}{N_1}(t=0).
\end{equation}
\\

\section*{III. Finite temperature reservoirs \label{app:nonzerotemp}}

\begin{figure*}\captionsetup{position=top}
\subfloat[]{\includegraphics[width=0.375\linewidth]{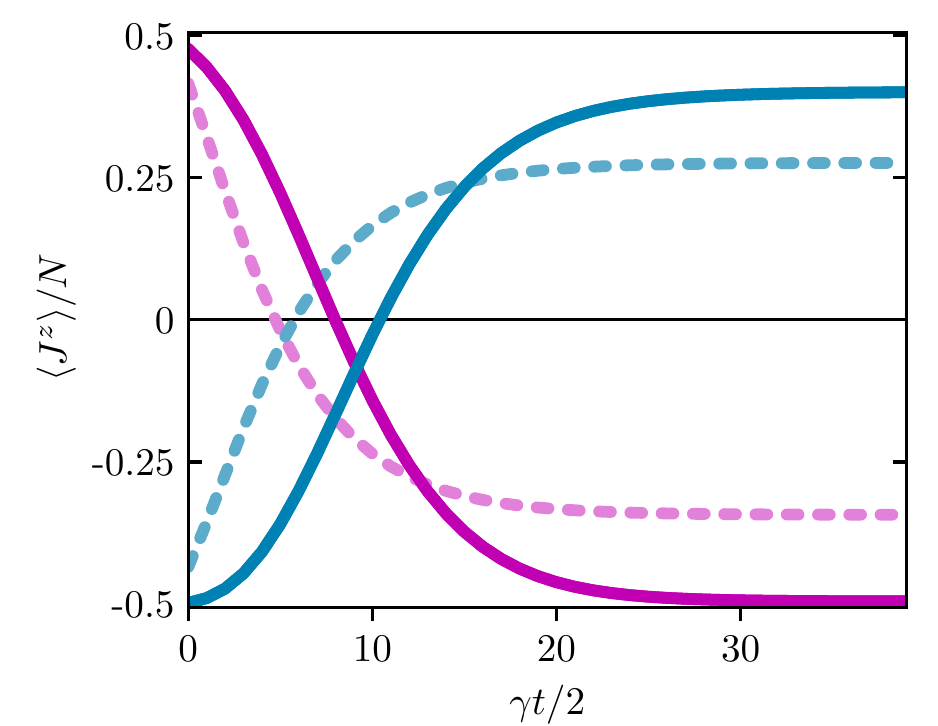}\label{fig:ft2}}
\subfloat[]{\includegraphics[width=0.3125\linewidth]{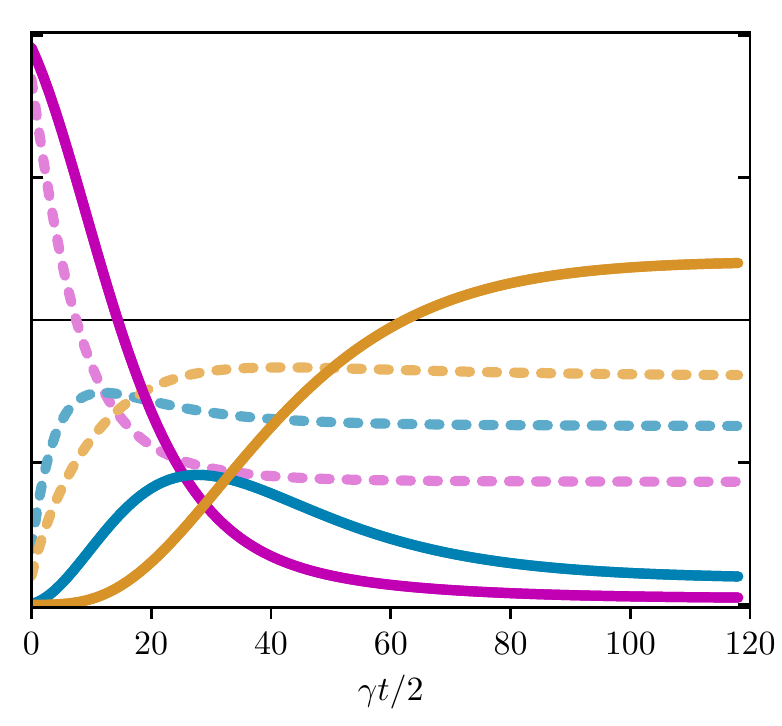}\label{fig:ft3}}
\subfloat[]{\includegraphics[width=0.3125\linewidth]{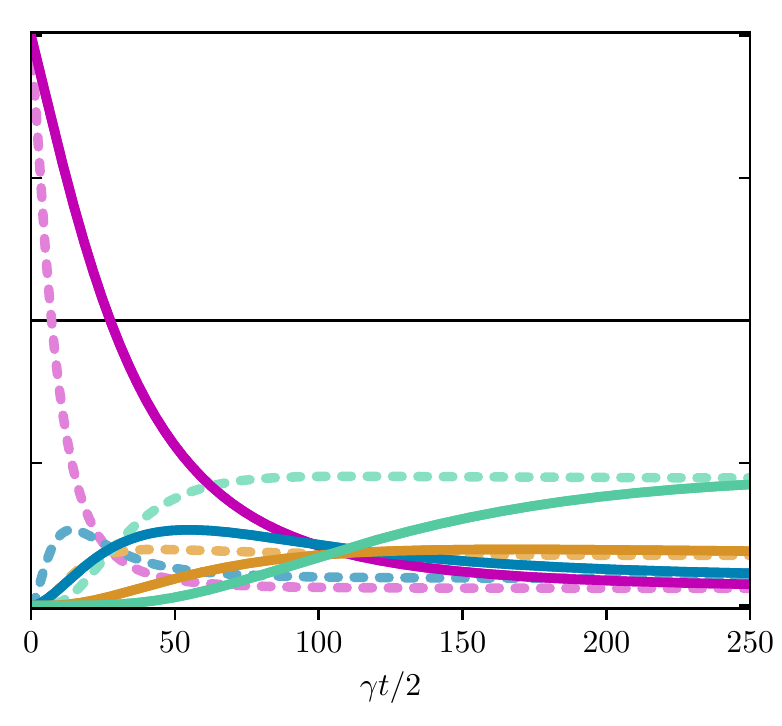}\label{fig:ft4}}\\
\subfloat{\includegraphics[width=0.9\linewidth]{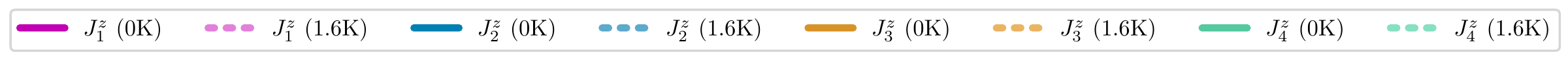}}
\caption{Collective spin relaxation dynamics with a zero temperature $T=0 K$  (solid curves) and finite temperature $T=1.6K$ reservoir (dashed lines)  \protect\subref{fig:ft2} corresponds to the two domain case with $N_1=20$ and $N_2=2$ while  \protect\subref{fig:ft3} corresponds to the three domain case with $N_1=12$, $N_2=6$ and $N_3=2$.  Finally \protect\subref{fig:ft4} corresponds to the four domain case with $N_1=4$, $N_2=3$, $N_3=3$ and $N_4=2$ respectively.  The following parameters are chosen: $\omega_0/2\pi=10 \mathrm{G Hz}$ and $\gamma_1=\gamma_2=0.01 \mathrm{Hz}$.}
\label{fig:ft}
\end{figure*}

While the main text explores the collective spin relaxation dynamics with zero temperature baths, in this section we explore the dynamics using baths at finite temperatures  Fig.~\ref{fig:ft} illustrates the dynamics for two, three and four domain systems (with $\omega_0/2\pi=10 \mathrm{G Hz}$ and $\gamma=0.01 \mathrm{Hz}$), showing simultaneously the collective spin relaxation for zero temperature baths (solid lines) and $T=1.6K$ baths (dashed lines). 

The two domain plot in Fig.~\ref{fig:ft}\subref{fig:ft2} clearly shows the steady state of  $J^z_2$ is less excited and the steady state of $J^z_1$ is more excited for non-zero temperature,  indicating a less efficient excitation transfer from the first domain to the second.  This effect is also shown in the three domain plot in Fig.~\ref{fig:ft}\subref{fig:ft3}.  Additionally,  both the two and three domain results, indicate a faster collective spin relaxation process at non-zero temperature.  
The four domain plot in Fig.~\ref{fig:ft}\subref{fig:ft4} shows the effects of thermalisation as in \ref{fig:ft}\subref{fig:ft2} and \ref{fig:ft}\subref{fig:ft3} albeit less pronounced due to the smaller system size (which is limited by the scaling of the Hilbert space as the number of domains is increased).


\begin{thebibliography}{49}%
\makeatletter
\providecommand \@ifxundefined [1]{%
 \@ifx{#1\undefined}
}%
\providecommand \@ifnum [1]{%
 \ifnum #1\expandafter \@firstoftwo
 \else \expandafter \@secondoftwo
 \fi
}%
\providecommand \@ifx [1]{%
 \ifx #1\expandafter \@firstoftwo
 \else \expandafter \@secondoftwo
 \fi
}%
\providecommand \natexlab [1]{#1}%
\providecommand \enquote  [1]{``#1''}%
\providecommand \bibnamefont  [1]{#1}%
\providecommand \bibfnamefont [1]{#1}%
\providecommand \citenamefont [1]{#1}%
\providecommand \href@noop [0]{\@secondoftwo}%
\providecommand \href [0]{\begingroup \@sanitize@url \@href}%
\providecommand \@href[1]{\@@startlink{#1}\@@href}%
\providecommand \@@href[1]{\endgroup#1\@@endlink}%
\providecommand \@sanitize@url [0]{\catcode `\\12\catcode `\$12\catcode
  `\&12\catcode `\#12\catcode `\^12\catcode `\_12\catcode `\%12\relax}%
\providecommand \@@startlink[1]{}%
\providecommand \@@endlink[0]{}%
\providecommand \url  [0]{\begingroup\@sanitize@url \@url }%
\providecommand \@url [1]{\endgroup\@href {#1}{\urlprefix }}%
\providecommand \urlprefix  [0]{URL }%
\providecommand \Eprint [0]{\href }%
\providecommand \doibase [0]{https://doi.org/}%
\providecommand \selectlanguage [0]{\@gobble}%
\providecommand \bibinfo  [0]{\@secondoftwo}%
\providecommand \bibfield  [0]{\@secondoftwo}%
\providecommand \translation [1]{[#1]}%
\providecommand \BibitemOpen [0]{}%
\providecommand \bibitemStop [0]{}%
\providecommand \bibitemNoStop [0]{.\EOS\space}%
\providecommand \EOS [0]{\spacefactor3000\relax}%
\providecommand \BibitemShut  [1]{\csname bibitem#1\endcsname}%
\let\auto@bib@innerbib\@empty
\bibitem [{\citenamefont {Krinner}\ \emph {et~al.}(2017)\citenamefont
  {Krinner}, \citenamefont {Esslinger},\ and\ \citenamefont
  {Brantut}}]{krinner2017Transport}%
  \BibitemOpen
  \bibfield  {author} {\bibinfo {author} {\bibfnamefont {S.}~\bibnamefont
  {Krinner}}, \bibinfo {author} {\bibfnamefont {T.}~\bibnamefont {Esslinger}},\
  and\ \bibinfo {author} {\bibfnamefont {J.-P.}\ \bibnamefont {Brantut}},\
  }\bibfield  {title} {\bibinfo {title} {Two-terminal transport measurements
  with cold atoms},\ }\href@noop {} {\bibfield  {journal} {\bibinfo  {journal}
  {J. Phys. Condens. Matter}\ }\textbf {\bibinfo {volume} {29}},\ \bibinfo
  {pages} {343003} (\bibinfo {year} {2017})}\BibitemShut {NoStop}%
\bibitem [{\citenamefont {Brown}\ \emph {et~al.}(2019)\citenamefont {Brown},
  \citenamefont {Mitra}, \citenamefont {Guardado-Sanchez}, \citenamefont
  {Nourafkan}, \citenamefont {Reymbaut}, \citenamefont {H{\'e}bert},
  \citenamefont {Bergeron}, \citenamefont {Tremblay}, \citenamefont {Kokalj},
  \citenamefont {Huse} \emph {et~al.}}]{brown2019bad}%
  \BibitemOpen
  \bibfield  {author} {\bibinfo {author} {\bibfnamefont {P.~T.}\ \bibnamefont
  {Brown}}, \bibinfo {author} {\bibfnamefont {D.}~\bibnamefont {Mitra}},
  \bibinfo {author} {\bibfnamefont {E.}~\bibnamefont {Guardado-Sanchez}},
  \bibinfo {author} {\bibfnamefont {R.}~\bibnamefont {Nourafkan}}, \bibinfo
  {author} {\bibfnamefont {A.}~\bibnamefont {Reymbaut}}, \bibinfo {author}
  {\bibfnamefont {C.-D.}\ \bibnamefont {H{\'e}bert}}, \bibinfo {author}
  {\bibfnamefont {S.}~\bibnamefont {Bergeron}}, \bibinfo {author}
  {\bibfnamefont {A.-M.}\ \bibnamefont {Tremblay}}, \bibinfo {author}
  {\bibfnamefont {J.}~\bibnamefont {Kokalj}}, \bibinfo {author} {\bibfnamefont
  {D.~A.}\ \bibnamefont {Huse}}, \emph {et~al.},\ }\bibfield  {title} {\bibinfo
  {title} {Bad metallic transport in a cold atom fermi-hubbard system},\
  }\href@noop {} {\bibfield  {journal} {\bibinfo  {journal} {Science}\ }\textbf
  {\bibinfo {volume} {363}},\ \bibinfo {pages} {379} (\bibinfo {year}
  {2019})}\BibitemShut {NoStop}%
\bibitem [{\citenamefont {Grohs}\ \emph {et~al.}(2016)\citenamefont {Grohs},
  \citenamefont {Fuller}, \citenamefont {Kishimoto}, \citenamefont {Paris},\
  and\ \citenamefont {Vlasenko}}]{Grohs2016}%
  \BibitemOpen
  \bibfield  {author} {\bibinfo {author} {\bibfnamefont {E.}~\bibnamefont
  {Grohs}}, \bibinfo {author} {\bibfnamefont {G.~M.}\ \bibnamefont {Fuller}},
  \bibinfo {author} {\bibfnamefont {C.~T.}\ \bibnamefont {Kishimoto}}, \bibinfo
  {author} {\bibfnamefont {M.~W.}\ \bibnamefont {Paris}},\ and\ \bibinfo
  {author} {\bibfnamefont {A.}~\bibnamefont {Vlasenko}},\ }\bibfield  {title}
  {\bibinfo {title} {Neutrino energy transport in weak decoupling and big bang
  nucleosynthesis},\ }\href {https://doi.org/10.1103/PhysRevD.93.083522}
  {\bibfield  {journal} {\bibinfo  {journal} {Phys. Rev. D}\ }\textbf {\bibinfo
  {volume} {93}},\ \bibinfo {pages} {083522} (\bibinfo {year}
  {2016})}\BibitemShut {NoStop}%
\bibitem [{\citenamefont {Szyd{\l}owski}(2006)}]{szydlowski2006cosmological}%
  \BibitemOpen
  \bibfield  {author} {\bibinfo {author} {\bibfnamefont {M.}~\bibnamefont
  {Szyd{\l}owski}},\ }\bibfield  {title} {\bibinfo {title} {Cosmological model
  with energy transfer},\ }\href@noop {} {\bibfield  {journal} {\bibinfo
  {journal} {Phys. Lett. B}\ }\textbf {\bibinfo {volume} {632}},\ \bibinfo
  {pages} {1} (\bibinfo {year} {2006})}\BibitemShut {NoStop}%
\bibitem [{\citenamefont {Collini}\ \emph {et~al.}(2010)\citenamefont
  {Collini}, \citenamefont {Wong}, \citenamefont {Wilk}, \citenamefont {Curmi},
  \citenamefont {Brumer},\ and\ \citenamefont
  {Scholes}}]{collini2010coherently}%
  \BibitemOpen
  \bibfield  {author} {\bibinfo {author} {\bibfnamefont {E.}~\bibnamefont
  {Collini}}, \bibinfo {author} {\bibfnamefont {C.~Y.}\ \bibnamefont {Wong}},
  \bibinfo {author} {\bibfnamefont {K.~E.}\ \bibnamefont {Wilk}}, \bibinfo
  {author} {\bibfnamefont {P.~M.~G.}\ \bibnamefont {Curmi}}, \bibinfo {author}
  {\bibfnamefont {P.}~\bibnamefont {Brumer}},\ and\ \bibinfo {author}
  {\bibfnamefont {G.~D.}\ \bibnamefont {Scholes}},\ }\bibfield  {title}
  {\bibinfo {title} {Coherently wired light-harvesting in photosynthetic marine
  algae at ambient temperature},\ }\href {https://doi.org/10.1038/nature08811}
  {\bibfield  {journal} {\bibinfo  {journal} {Nature}\ }\textbf {\bibinfo
  {volume} {463}},\ \bibinfo {pages} {644} (\bibinfo {year}
  {2010})}\BibitemShut {NoStop}%
\bibitem [{\citenamefont {Cao}\ \emph {et~al.}(2020)\citenamefont {Cao},
  \citenamefont {Cogdell}, \citenamefont {Coker}, \citenamefont {Duan},
  \citenamefont {Hauer}, \citenamefont {Kleinekath{\"o}fer}, \citenamefont
  {Jansen}, \citenamefont {Man{\v{c}}al}, \citenamefont {Miller}, \citenamefont
  {Ogilvie} \emph {et~al.}}]{cao2020quantum}%
  \BibitemOpen
  \bibfield  {author} {\bibinfo {author} {\bibfnamefont {J.}~\bibnamefont
  {Cao}}, \bibinfo {author} {\bibfnamefont {R.~J.}\ \bibnamefont {Cogdell}},
  \bibinfo {author} {\bibfnamefont {D.~F.}\ \bibnamefont {Coker}}, \bibinfo
  {author} {\bibfnamefont {H.-G.}\ \bibnamefont {Duan}}, \bibinfo {author}
  {\bibfnamefont {J.}~\bibnamefont {Hauer}}, \bibinfo {author} {\bibfnamefont
  {U.}~\bibnamefont {Kleinekath{\"o}fer}}, \bibinfo {author} {\bibfnamefont
  {T.~L.}\ \bibnamefont {Jansen}}, \bibinfo {author} {\bibfnamefont
  {T.}~\bibnamefont {Man{\v{c}}al}}, \bibinfo {author} {\bibfnamefont {R.~D.}\
  \bibnamefont {Miller}}, \bibinfo {author} {\bibfnamefont {J.~P.}\
  \bibnamefont {Ogilvie}}, \emph {et~al.},\ }\bibfield  {title} {\bibinfo
  {title} {Quantum biology revisited},\ }\href
  {https://doi.org/10.1126/sciadv.aaz4888} {\bibfield  {journal} {\bibinfo
  {journal} {Sci. Adv.}\ }\textbf {\bibinfo {volume} {6}},\ \bibinfo {pages}
  {eaaz4888} (\bibinfo {year} {2020})}\BibitemShut {NoStop}%
\bibitem [{\citenamefont {Lambert}\ \emph {et~al.}(2013)\citenamefont
  {Lambert}, \citenamefont {Chen}, \citenamefont {Cheng}, \citenamefont {Li},
  \citenamefont {Chen},\ and\ \citenamefont {Nori}}]{lambert2013quantum}%
  \BibitemOpen
  \bibfield  {author} {\bibinfo {author} {\bibfnamefont {N.}~\bibnamefont
  {Lambert}}, \bibinfo {author} {\bibfnamefont {Y.-N.}\ \bibnamefont {Chen}},
  \bibinfo {author} {\bibfnamefont {Y.-C.}\ \bibnamefont {Cheng}}, \bibinfo
  {author} {\bibfnamefont {C.-M.}\ \bibnamefont {Li}}, \bibinfo {author}
  {\bibfnamefont {G.-Y.}\ \bibnamefont {Chen}},\ and\ \bibinfo {author}
  {\bibfnamefont {F.}~\bibnamefont {Nori}},\ }\bibfield  {title} {\bibinfo
  {title} {Quantum biology},\ }\href {https://doi.org/10.1038/nphys2474}
  {\bibfield  {journal} {\bibinfo  {journal} {Nat. Phys.}\ }\textbf {\bibinfo
  {volume} {9}},\ \bibinfo {pages} {10} (\bibinfo {year} {2013})}\BibitemShut
  {NoStop}%
\bibitem [{\citenamefont {Kashida}\ \emph {et~al.}(2018)\citenamefont
  {Kashida}, \citenamefont {Kawai}, \citenamefont {Maruyama}, \citenamefont
  {Kokubo}, \citenamefont {Araki}, \citenamefont {Wada},\ and\ \citenamefont
  {Asanuma}}]{kashida2018quantitative}%
  \BibitemOpen
  \bibfield  {author} {\bibinfo {author} {\bibfnamefont {H.}~\bibnamefont
  {Kashida}}, \bibinfo {author} {\bibfnamefont {H.}~\bibnamefont {Kawai}},
  \bibinfo {author} {\bibfnamefont {R.}~\bibnamefont {Maruyama}}, \bibinfo
  {author} {\bibfnamefont {Y.}~\bibnamefont {Kokubo}}, \bibinfo {author}
  {\bibfnamefont {Y.}~\bibnamefont {Araki}}, \bibinfo {author} {\bibfnamefont
  {T.}~\bibnamefont {Wada}},\ and\ \bibinfo {author} {\bibfnamefont
  {H.}~\bibnamefont {Asanuma}},\ }\bibfield  {title} {\bibinfo {title}
  {Quantitative evaluation of energy migration between identical chromophores
  enabled by breaking symmetry},\ }\href
  {https://doi.org/10.1038/s42004-018-0093-0} {\bibfield  {journal} {\bibinfo
  {journal} {Commun. Chem.}\ }\textbf {\bibinfo {volume} {1}},\ \bibinfo
  {pages} {91} (\bibinfo {year} {2018})}\BibitemShut {NoStop}%
\bibitem [{\citenamefont {Franck}\ and\ \citenamefont
  {Teller}(1938)}]{franck1938migration}%
  \BibitemOpen
  \bibfield  {author} {\bibinfo {author} {\bibfnamefont {J.}~\bibnamefont
  {Franck}}\ and\ \bibinfo {author} {\bibfnamefont {E.}~\bibnamefont
  {Teller}},\ }\bibfield  {title} {\bibinfo {title} {Migration and
  photochemical action of excitation energy in crystals},\ }\href
  {https://doi.org/10.1063/1.1750182} {\bibfield  {journal} {\bibinfo
  {journal} {J. Chem. Phys.}\ }\textbf {\bibinfo {volume} {6}},\ \bibinfo
  {pages} {861} (\bibinfo {year} {1938})}\BibitemShut {NoStop}%
\bibitem [{\citenamefont {Brixner}\ \emph {et~al.}(2005)\citenamefont
  {Brixner}, \citenamefont {Stenger}, \citenamefont {Vaswani}, \citenamefont
  {Cho}, \citenamefont {Blankenship},\ and\ \citenamefont
  {Fleming}}]{brixner2005two}%
  \BibitemOpen
  \bibfield  {author} {\bibinfo {author} {\bibfnamefont {T.}~\bibnamefont
  {Brixner}}, \bibinfo {author} {\bibfnamefont {J.}~\bibnamefont {Stenger}},
  \bibinfo {author} {\bibfnamefont {H.~M.}\ \bibnamefont {Vaswani}}, \bibinfo
  {author} {\bibfnamefont {M.}~\bibnamefont {Cho}}, \bibinfo {author}
  {\bibfnamefont {R.~E.}\ \bibnamefont {Blankenship}},\ and\ \bibinfo {author}
  {\bibfnamefont {G.~R.}\ \bibnamefont {Fleming}},\ }\bibfield  {title}
  {\bibinfo {title} {Two-dimensional spectroscopy of electronic couplings in
  photosynthesis},\ }\href {https://doi.org/10.1038/nature03429} {\bibfield
  {journal} {\bibinfo  {journal} {Nature}\ }\textbf {\bibinfo {volume} {434}},\
  \bibinfo {pages} {625} (\bibinfo {year} {2005})}\BibitemShut {NoStop}%
\bibitem [{\citenamefont {Christensson}\ \emph {et~al.}(2012)\citenamefont
  {Christensson}, \citenamefont {Kauffmann}, \citenamefont {Pullerits},\ and\
  \citenamefont {Mančal}}]{christensson2012origin}%
  \BibitemOpen
  \bibfield  {author} {\bibinfo {author} {\bibfnamefont {N.}~\bibnamefont
  {Christensson}}, \bibinfo {author} {\bibfnamefont {H.~F.}\ \bibnamefont
  {Kauffmann}}, \bibinfo {author} {\bibfnamefont {T.}~\bibnamefont
  {Pullerits}},\ and\ \bibinfo {author} {\bibfnamefont {T.}~\bibnamefont
  {Mančal}},\ }\bibfield  {title} {\bibinfo {title} {Origin of long-lived
  coherences in light-harvesting complexes},\ }\href
  {https://doi.org/10.1021/jp304649c} {\bibfield  {journal} {\bibinfo
  {journal} {J. Phys. Chem. B}\ }\textbf {\bibinfo {volume} {116}},\ \bibinfo
  {pages} {7449} (\bibinfo {year} {2012})}\BibitemShut {NoStop}%
\bibitem [{\citenamefont {Ritschel}\ \emph {et~al.}(2011)\citenamefont
  {Ritschel}, \citenamefont {Roden}, \citenamefont {Strunz},\ and\
  \citenamefont {Eisfeld}}]{ritschel2011efficient}%
  \BibitemOpen
  \bibfield  {author} {\bibinfo {author} {\bibfnamefont {G.}~\bibnamefont
  {Ritschel}}, \bibinfo {author} {\bibfnamefont {J.}~\bibnamefont {Roden}},
  \bibinfo {author} {\bibfnamefont {W.~T.}\ \bibnamefont {Strunz}},\ and\
  \bibinfo {author} {\bibfnamefont {A.}~\bibnamefont {Eisfeld}},\ }\bibfield
  {title} {\bibinfo {title} {An efficient method to calculate excitation energy
  transfer in light-harvesting systems: application to the
  fenna{\textendash}matthews{\textendash}olson complex},\ }\href
  {https://doi.org/10.1088/1367-2630/13/11/113034} {\bibfield  {journal}
  {\bibinfo  {journal} {New J. Phys.}\ }\textbf {\bibinfo {volume} {13}},\
  \bibinfo {pages} {113034} (\bibinfo {year} {2011})}\BibitemShut {NoStop}%
\bibitem [{\citenamefont {Briggs}\ and\ \citenamefont
  {Eisfeld}(2011)}]{briggs2011equivalence}%
  \BibitemOpen
  \bibfield  {author} {\bibinfo {author} {\bibfnamefont {J.~S.}\ \bibnamefont
  {Briggs}}\ and\ \bibinfo {author} {\bibfnamefont {A.}~\bibnamefont
  {Eisfeld}},\ }\bibfield  {title} {\bibinfo {title} {Equivalence of quantum
  and classical coherence in electronic energy transfer},\ }\href
  {https://doi.org/10.1103/PhysRevE.83.051911} {\bibfield  {journal} {\bibinfo
  {journal} {Phys. Rev. E}\ }\textbf {\bibinfo {volume} {83}},\ \bibinfo
  {pages} {051911} (\bibinfo {year} {2011})}\BibitemShut {NoStop}%
\bibitem [{\citenamefont {Breuer}\ and\ \citenamefont
  {Petruccione}(2002)}]{BreuerPetruccioneBook2002}%
  \BibitemOpen
  \bibfield  {author} {\bibinfo {author} {\bibfnamefont {H.~P.}\ \bibnamefont
  {Breuer}}\ and\ \bibinfo {author} {\bibfnamefont {F.}~\bibnamefont
  {Petruccione}},\ }\href@noop {} {\emph {\bibinfo {title} {The Theory of Open
  Quantum Systems}}}\ (\bibinfo  {publisher} {Oxford University Press},\
  \bibinfo {address} {Oxford},\ \bibinfo {year} {2002})\BibitemShut {NoStop}%
\bibitem [{\citenamefont {Rebentrost}\ \emph {et~al.}(2009)\citenamefont
  {Rebentrost}, \citenamefont {Mohseni}, \citenamefont {Kassal}, \citenamefont
  {Lloyd},\ and\ \citenamefont {Aspuru-Guzik}}]{rebentrost2009environment}%
  \BibitemOpen
  \bibfield  {author} {\bibinfo {author} {\bibfnamefont {P.}~\bibnamefont
  {Rebentrost}}, \bibinfo {author} {\bibfnamefont {M.}~\bibnamefont {Mohseni}},
  \bibinfo {author} {\bibfnamefont {I.}~\bibnamefont {Kassal}}, \bibinfo
  {author} {\bibfnamefont {S.}~\bibnamefont {Lloyd}},\ and\ \bibinfo {author}
  {\bibfnamefont {A.}~\bibnamefont {Aspuru-Guzik}},\ }\bibfield  {title}
  {\bibinfo {title} {Environment-assisted quantum transport},\ }\href
  {https://doi.org/10.1088/1367-2630/11/3/033003} {\bibfield  {journal}
  {\bibinfo  {journal} {New J. Phys.}\ }\textbf {\bibinfo {volume} {11}},\
  \bibinfo {pages} {033003} (\bibinfo {year} {2009})}\BibitemShut {NoStop}%
\bibitem [{\citenamefont {Plenio}\ and\ \citenamefont
  {Huelga}(2008)}]{plenio2008dephasing}%
  \BibitemOpen
  \bibfield  {author} {\bibinfo {author} {\bibfnamefont {M.~B.}\ \bibnamefont
  {Plenio}}\ and\ \bibinfo {author} {\bibfnamefont {S.~F.}\ \bibnamefont
  {Huelga}},\ }\bibfield  {title} {\bibinfo {title} {Dephasing-assisted
  transport: quantum networks and biomolecules},\ }\href
  {https://doi.org/10.1088/1367-2630/10/11/113019} {\bibfield  {journal}
  {\bibinfo  {journal} {New J. Phys.}\ }\textbf {\bibinfo {volume} {10}},\
  \bibinfo {pages} {113019} (\bibinfo {year} {2008})}\BibitemShut {NoStop}%
\bibitem [{\citenamefont {Gaab}\ and\ \citenamefont
  {Bardeen}(2004)}]{gaab2004effects}%
  \BibitemOpen
  \bibfield  {author} {\bibinfo {author} {\bibfnamefont {K.~M.}\ \bibnamefont
  {Gaab}}\ and\ \bibinfo {author} {\bibfnamefont {C.~J.}\ \bibnamefont
  {Bardeen}},\ }\bibfield  {title} {\bibinfo {title} {The effects of
  connectivity, coherence, and trapping on energy transfer in simple
  light-harvesting systems studied using the haken-strobl model with diagonal
  disorder},\ }\href {https://doi.org/10.1063/1.1786922} {\bibfield  {journal}
  {\bibinfo  {journal} {J. Chem. Phys}\ }\textbf {\bibinfo {volume} {121}},\
  \bibinfo {pages} {7813} (\bibinfo {year} {2004})}\BibitemShut {NoStop}%
\bibitem [{\citenamefont {Biggerstaff}\ \emph {et~al.}(2016)\citenamefont
  {Biggerstaff}, \citenamefont {Heilmann}, \citenamefont {Zecevik},
  \citenamefont {Gräfe}, \citenamefont {Broome}, \citenamefont {Fedrizzi},
  \citenamefont {Nolte}, \citenamefont {Szameit}, \citenamefont {White},\ and\
  \citenamefont {Kassal}}]{biggerstaff2016enhancing}%
  \BibitemOpen
  \bibfield  {author} {\bibinfo {author} {\bibfnamefont {D.~N.}\ \bibnamefont
  {Biggerstaff}}, \bibinfo {author} {\bibfnamefont {R.}~\bibnamefont
  {Heilmann}}, \bibinfo {author} {\bibfnamefont {A.~A.}\ \bibnamefont
  {Zecevik}}, \bibinfo {author} {\bibfnamefont {M.}~\bibnamefont {Gräfe}},
  \bibinfo {author} {\bibfnamefont {M.~A.}\ \bibnamefont {Broome}}, \bibinfo
  {author} {\bibfnamefont {A.}~\bibnamefont {Fedrizzi}}, \bibinfo {author}
  {\bibfnamefont {S.}~\bibnamefont {Nolte}}, \bibinfo {author} {\bibfnamefont
  {A.}~\bibnamefont {Szameit}}, \bibinfo {author} {\bibfnamefont {A.~G.}\
  \bibnamefont {White}},\ and\ \bibinfo {author} {\bibfnamefont
  {I.}~\bibnamefont {Kassal}},\ }\bibfield  {title} {\bibinfo {title}
  {Enhancing coherent transport in a photonic network using controllable
  decoherence},\ }\href {https://doi.org/10.1038/ncomms11282} {\bibfield
  {journal} {\bibinfo  {journal} {Nat. Commun.}\ }\textbf {\bibinfo {volume}
  {7}},\ \bibinfo {pages} {11282} (\bibinfo {year} {2016})}\BibitemShut
  {NoStop}%
\bibitem [{\citenamefont {Uchiyama}\ \emph {et~al.}(2018)\citenamefont
  {Uchiyama}, \citenamefont {Munro},\ and\ \citenamefont
  {Nemoto}}]{uchiyama2018environmental}%
  \BibitemOpen
  \bibfield  {author} {\bibinfo {author} {\bibfnamefont {C.}~\bibnamefont
  {Uchiyama}}, \bibinfo {author} {\bibfnamefont {W.~J.}\ \bibnamefont
  {Munro}},\ and\ \bibinfo {author} {\bibfnamefont {K.}~\bibnamefont
  {Nemoto}},\ }\bibfield  {title} {\bibinfo {title} {Environmental engineering
  for quantum energy transport},\ }\href
  {https://doi.org/10.1038/s41534-018-0079-x} {\bibfield  {journal} {\bibinfo
  {journal} {npj Quantum Inf.}\ }\textbf {\bibinfo {volume} {4}},\ \bibinfo
  {pages} {1} (\bibinfo {year} {2018})}\BibitemShut {NoStop}%
\bibitem [{\citenamefont {Zhang}\ \emph {et~al.}(2017)\citenamefont {Zhang},
  \citenamefont {Celardo}, \citenamefont {Borgonovi},\ and\ \citenamefont
  {Kaplan}}]{Zhang2017}%
  \BibitemOpen
  \bibfield  {author} {\bibinfo {author} {\bibfnamefont {Y.}~\bibnamefont
  {Zhang}}, \bibinfo {author} {\bibfnamefont {G.~L.}\ \bibnamefont {Celardo}},
  \bibinfo {author} {\bibfnamefont {F.}~\bibnamefont {Borgonovi}},\ and\
  \bibinfo {author} {\bibfnamefont {L.}~\bibnamefont {Kaplan}},\ }\bibfield
  {title} {\bibinfo {title} {Optimal dephasing for ballistic energy transfer in
  disordered linear chains},\ }\href
  {https://doi.org/10.1103/PhysRevE.96.052103} {\bibfield  {journal} {\bibinfo
  {journal} {Phys. Rev. E}\ }\textbf {\bibinfo {volume} {96}},\ \bibinfo
  {pages} {052103} (\bibinfo {year} {2017})}\BibitemShut {NoStop}%
\bibitem [{\citenamefont {Einstein}\ \emph {et~al.}(1935)\citenamefont
  {Einstein}, \citenamefont {Podolsky},\ and\ \citenamefont
  {Rosen}}]{einstein1935can}%
  \BibitemOpen
  \bibfield  {author} {\bibinfo {author} {\bibfnamefont {A.}~\bibnamefont
  {Einstein}}, \bibinfo {author} {\bibfnamefont {B.}~\bibnamefont {Podolsky}},\
  and\ \bibinfo {author} {\bibfnamefont {N.}~\bibnamefont {Rosen}},\ }\bibfield
   {title} {\bibinfo {title} {Can quantum-mechanical description of physical
  reality be considered complete?},\ }\href
  {https://doi.org/10.1103/PhysRev.47.777} {\bibfield  {journal} {\bibinfo
  {journal} {Phys. Rev.}\ }\textbf {\bibinfo {volume} {47}},\ \bibinfo {pages}
  {777} (\bibinfo {year} {1935})}\BibitemShut {NoStop}%
\bibitem [{\citenamefont {Peres}(2006)}]{peres2006quantum}%
  \BibitemOpen
  \bibfield  {author} {\bibinfo {author} {\bibfnamefont {A.}~\bibnamefont
  {Peres}},\ }\href@noop {} {\emph {\bibinfo {title} {Quantum theory: concepts
  and methods}}},\ Vol.~\bibinfo {volume} {57}\ (\bibinfo  {publisher}
  {Springer Science \& Business Media},\ \bibinfo {year} {2006})\BibitemShut
  {NoStop}%
\bibitem [{\citenamefont {Matsuzaki}\ \emph {et~al.}(2020)\citenamefont
  {Matsuzaki}, \citenamefont {Bastidas}, \citenamefont {Takeuchi},
  \citenamefont {Munro},\ and\ \citenamefont {Saito}}]{Matsuzaki2020}%
  \BibitemOpen
  \bibfield  {author} {\bibinfo {author} {\bibfnamefont {Y.}~\bibnamefont
  {Matsuzaki}}, \bibinfo {author} {\bibfnamefont {V.~M.}\ \bibnamefont
  {Bastidas}}, \bibinfo {author} {\bibfnamefont {Y.}~\bibnamefont {Takeuchi}},
  \bibinfo {author} {\bibfnamefont {W.~J.}\ \bibnamefont {Munro}},\ and\
  \bibinfo {author} {\bibfnamefont {S.}~\bibnamefont {Saito}},\ }\bibfield
  {title} {\bibinfo {title} {One-way transfer of quantum states via
  decoherence},\ }\href {https://doi.org/10.7566/JPSJ.89.044003} {\bibfield
  {journal} {\bibinfo  {journal} {J. Phys. Soc. Jpn.}\ }\textbf {\bibinfo
  {volume} {89}},\ \bibinfo {pages} {044003} (\bibinfo {year}
  {2020})}\BibitemShut {NoStop}%
\bibitem [{\citenamefont {Lieb}\ and\ \citenamefont
  {Robinson}(1972)}]{Lieb1972}%
  \BibitemOpen
  \bibfield  {author} {\bibinfo {author} {\bibfnamefont {E.~H.}\ \bibnamefont
  {Lieb}}\ and\ \bibinfo {author} {\bibfnamefont {D.~W.}\ \bibnamefont
  {Robinson}},\ }\bibfield  {title} {\bibinfo {title} {The finite group
  velocity of quantum spin systems},\ }\href
  {https://doi.org/10.1007/BF01645779} {\bibfield  {journal} {\bibinfo
  {journal} {Commun. Math. Phys.}\ }\textbf {\bibinfo {volume} {28}},\ \bibinfo
  {pages} {251} (\bibinfo {year} {1972})}\BibitemShut {NoStop}%
\bibitem [{\citenamefont {Deffner}\ and\ \citenamefont
  {Campbell}(2017)}]{deffner2017quantum}%
  \BibitemOpen
  \bibfield  {author} {\bibinfo {author} {\bibfnamefont {S.}~\bibnamefont
  {Deffner}}\ and\ \bibinfo {author} {\bibfnamefont {S.}~\bibnamefont
  {Campbell}},\ }\bibfield  {title} {\bibinfo {title} {Quantum speed limits:
  from heisenberg's uncertainty principle to optimal quantum control},\ }\href
  {https://doi.org/10.1088/1751-8121/aa86c6} {\bibfield  {journal} {\bibinfo
  {journal} {J. Phys. A}\ }\textbf {\bibinfo {volume} {50}},\ \bibinfo {pages}
  {453001} (\bibinfo {year} {2017})}\BibitemShut {NoStop}%
\bibitem [{\citenamefont {Tran}\ \emph {et~al.}(2019)\citenamefont {Tran},
  \citenamefont {Guo}, \citenamefont {Su}, \citenamefont {Garrison},
  \citenamefont {Eldredge}, \citenamefont {Foss-Feig}, \citenamefont {Childs},\
  and\ \citenamefont {Gorshkov}}]{Tran2019}%
  \BibitemOpen
  \bibfield  {author} {\bibinfo {author} {\bibfnamefont {M.~C.}\ \bibnamefont
  {Tran}}, \bibinfo {author} {\bibfnamefont {A.~Y.}\ \bibnamefont {Guo}},
  \bibinfo {author} {\bibfnamefont {Y.}~\bibnamefont {Su}}, \bibinfo {author}
  {\bibfnamefont {J.~R.}\ \bibnamefont {Garrison}}, \bibinfo {author}
  {\bibfnamefont {Z.}~\bibnamefont {Eldredge}}, \bibinfo {author}
  {\bibfnamefont {M.}~\bibnamefont {Foss-Feig}}, \bibinfo {author}
  {\bibfnamefont {A.~M.}\ \bibnamefont {Childs}},\ and\ \bibinfo {author}
  {\bibfnamefont {A.~V.}\ \bibnamefont {Gorshkov}},\ }\bibfield  {title}
  {\bibinfo {title} {Locality and digital quantum simulation of power-law
  interactions},\ }\href {https://doi.org/10.1103/PhysRevX.9.031006} {\bibfield
   {journal} {\bibinfo  {journal} {Phys. Rev. X}\ }\textbf {\bibinfo {volume}
  {9}},\ \bibinfo {pages} {031006} (\bibinfo {year} {2019})}\BibitemShut
  {NoStop}%
\bibitem [{\citenamefont {Verstraete}\ \emph {et~al.}(2009)\citenamefont
  {Verstraete}, \citenamefont {Wolf},\ and\ \citenamefont
  {Cirac}}]{verstraete2009quantum}%
  \BibitemOpen
  \bibfield  {author} {\bibinfo {author} {\bibfnamefont {F.}~\bibnamefont
  {Verstraete}}, \bibinfo {author} {\bibfnamefont {M.~M.}\ \bibnamefont
  {Wolf}},\ and\ \bibinfo {author} {\bibfnamefont {J.~I.}\ \bibnamefont
  {Cirac}},\ }\bibfield  {title} {\bibinfo {title} {Quantum computation and
  quantum-state engineering driven by dissipation},\ }\href
  {https://doi.org/10.1038/nphys1342} {\bibfield  {journal} {\bibinfo
  {journal} {Nat. Phys.}\ }\textbf {\bibinfo {volume} {5}},\ \bibinfo {pages}
  {633} (\bibinfo {year} {2009})}\BibitemShut {NoStop}%
\bibitem [{\citenamefont {Barreiro}\ \emph {et~al.}(2011)\citenamefont
  {Barreiro}, \citenamefont {M{\"u}ller}, \citenamefont {Schindler},
  \citenamefont {Nigg}, \citenamefont {Monz}, \citenamefont {Chwalla},
  \citenamefont {Hennrich}, \citenamefont {Roos}, \citenamefont {Zoller},\ and\
  \citenamefont {Blatt}}]{barreiro2011open}%
  \BibitemOpen
  \bibfield  {author} {\bibinfo {author} {\bibfnamefont {J.~T.}\ \bibnamefont
  {Barreiro}}, \bibinfo {author} {\bibfnamefont {M.}~\bibnamefont
  {M{\"u}ller}}, \bibinfo {author} {\bibfnamefont {P.}~\bibnamefont
  {Schindler}}, \bibinfo {author} {\bibfnamefont {D.}~\bibnamefont {Nigg}},
  \bibinfo {author} {\bibfnamefont {T.}~\bibnamefont {Monz}}, \bibinfo {author}
  {\bibfnamefont {M.}~\bibnamefont {Chwalla}}, \bibinfo {author} {\bibfnamefont
  {M.}~\bibnamefont {Hennrich}}, \bibinfo {author} {\bibfnamefont {C.~F.}\
  \bibnamefont {Roos}}, \bibinfo {author} {\bibfnamefont {P.}~\bibnamefont
  {Zoller}},\ and\ \bibinfo {author} {\bibfnamefont {R.}~\bibnamefont
  {Blatt}},\ }\bibfield  {title} {\bibinfo {title} {An open-system quantum
  simulator with trapped ions},\ }\href {https://doi.org/10.1038/nature09801}
  {\bibfield  {journal} {\bibinfo  {journal} {Nature}\ }\textbf {\bibinfo
  {volume} {470}},\ \bibinfo {pages} {486} (\bibinfo {year}
  {2011})}\BibitemShut {NoStop}%
\bibitem [{\citenamefont {Liu}\ and\ \citenamefont {Segal}(2020)}]{Liu2020}%
  \BibitemOpen
  \bibfield  {author} {\bibinfo {author} {\bibfnamefont {J.}~\bibnamefont
  {Liu}}\ and\ \bibinfo {author} {\bibfnamefont {D.}~\bibnamefont {Segal}},\
  }\bibfield  {title} {\bibinfo {title} {Dissipation engineering of
  nonreciprocal quantum dot circuits: An input-output approach},\ }\href
  {https://doi.org/10.1103/PhysRevB.102.125416} {\bibfield  {journal} {\bibinfo
   {journal} {Phys. Rev. B}\ }\textbf {\bibinfo {volume} {102}},\ \bibinfo
  {pages} {125416} (\bibinfo {year} {2020})}\BibitemShut {NoStop}%
\bibitem [{\citenamefont {Seetharam}\ \emph {et~al.}(2021)\citenamefont
  {Seetharam}, \citenamefont {Lerose}, \citenamefont {Fazio},\ and\
  \citenamefont {Marino}}]{seetharam2021}%
  \BibitemOpen
  \bibfield  {author} {\bibinfo {author} {\bibfnamefont {K.}~\bibnamefont
  {Seetharam}}, \bibinfo {author} {\bibfnamefont {A.}~\bibnamefont {Lerose}},
  \bibinfo {author} {\bibfnamefont {R.}~\bibnamefont {Fazio}},\ and\ \bibinfo
  {author} {\bibfnamefont {J.}~\bibnamefont {Marino}},\ }\bibfield  {title}
  {\bibinfo {title} {Correlation engineering via non-local dissipation},\
  }\href@noop {} {\bibfield  {journal} {\bibinfo  {journal} {arXiv:2101.06445}\
  } (\bibinfo {year} {2021})}\BibitemShut {NoStop}%
\bibitem [{\citenamefont {Angerer}\ \emph {et~al.}(2018)\citenamefont
  {Angerer}, \citenamefont {Streltsov}, \citenamefont {Astner}, \citenamefont
  {Putz}, \citenamefont {Sumiya}, \citenamefont {Onoda}, \citenamefont {Isoya},
  \citenamefont {Munro}, \citenamefont {Nemoto}, \citenamefont {Schmiedmayer},\
  and\ \citenamefont {Majer}}]{angerer2018superradiant}%
  \BibitemOpen
  \bibfield  {author} {\bibinfo {author} {\bibfnamefont {A.}~\bibnamefont
  {Angerer}}, \bibinfo {author} {\bibfnamefont {K.}~\bibnamefont {Streltsov}},
  \bibinfo {author} {\bibfnamefont {T.}~\bibnamefont {Astner}}, \bibinfo
  {author} {\bibfnamefont {S.}~\bibnamefont {Putz}}, \bibinfo {author}
  {\bibfnamefont {H.}~\bibnamefont {Sumiya}}, \bibinfo {author} {\bibfnamefont
  {S.}~\bibnamefont {Onoda}}, \bibinfo {author} {\bibfnamefont
  {J.}~\bibnamefont {Isoya}}, \bibinfo {author} {\bibfnamefont {W.~J.}\
  \bibnamefont {Munro}}, \bibinfo {author} {\bibfnamefont {K.}~\bibnamefont
  {Nemoto}}, \bibinfo {author} {\bibfnamefont {J.}~\bibnamefont
  {Schmiedmayer}},\ and\ \bibinfo {author} {\bibfnamefont {J.}~\bibnamefont
  {Majer}},\ }\bibfield  {title} {\bibinfo {title} {Superradiant emission from
  colour centres in diamond},\ }\href
  {https://doi.org/10.1038/s41567-018-0269-7} {\bibfield  {journal} {\bibinfo
  {journal} {Nat. Phys.}\ }\textbf {\bibinfo {volume} {14}},\ \bibinfo {pages}
  {1168} (\bibinfo {year} {2018})}\BibitemShut {NoStop}%
\bibitem [{\citenamefont {Dicke}(1954)}]{dicke1954coherence}%
  \BibitemOpen
  \bibfield  {author} {\bibinfo {author} {\bibfnamefont {R.~H.}\ \bibnamefont
  {Dicke}},\ }\bibfield  {title} {\bibinfo {title} {Coherence in spontaneous
  radiation processes},\ }\href {https://doi.org/10.1103/PhysRev.93.99}
  {\bibfield  {journal} {\bibinfo  {journal} {Phys. Rev.}\ }\textbf {\bibinfo
  {volume} {93}},\ \bibinfo {pages} {99} (\bibinfo {year} {1954})}\BibitemShut
  {NoStop}%
\bibitem [{\citenamefont {Gross}\ and\ \citenamefont
  {Haroche}(1982)}]{gross1982superradiance}%
  \BibitemOpen
  \bibfield  {author} {\bibinfo {author} {\bibfnamefont {M.}~\bibnamefont
  {Gross}}\ and\ \bibinfo {author} {\bibfnamefont {S.}~\bibnamefont
  {Haroche}},\ }\bibfield  {title} {\bibinfo {title} {Superradiance: An essay
  on the theory of collective spontaneous emission},\ }\href
  {https://doi.org/https://doi.org/10.1016/0370-1573(82)90102-8} {\bibfield
  {journal} {\bibinfo  {journal} {Phys. Rep.}\ }\textbf {\bibinfo {volume}
  {93}},\ \bibinfo {pages} {301} (\bibinfo {year} {1982})}\BibitemShut
  {NoStop}%
\bibitem [{\citenamefont {Yang}\ \emph {et~al.}(2021)\citenamefont {Yang},
  \citenamefont {Oh}, \citenamefont {Han}, \citenamefont {Son}, \citenamefont
  {Kim}, \citenamefont {Kim}, \citenamefont {Lee},\ and\ \citenamefont
  {An}}]{yang2021realization}%
  \BibitemOpen
  \bibfield  {author} {\bibinfo {author} {\bibfnamefont {D.}~\bibnamefont
  {Yang}}, \bibinfo {author} {\bibfnamefont {S.-h.}\ \bibnamefont {Oh}},
  \bibinfo {author} {\bibfnamefont {J.}~\bibnamefont {Han}}, \bibinfo {author}
  {\bibfnamefont {G.}~\bibnamefont {Son}}, \bibinfo {author} {\bibfnamefont
  {J.}~\bibnamefont {Kim}}, \bibinfo {author} {\bibfnamefont {J.}~\bibnamefont
  {Kim}}, \bibinfo {author} {\bibfnamefont {M.}~\bibnamefont {Lee}},\ and\
  \bibinfo {author} {\bibfnamefont {K.}~\bibnamefont {An}},\ }\bibfield
  {title} {\bibinfo {title} {Realization of superabsorption by time reversal of
  superradiance},\ }\href {https://doi.org/10.1038/s41566-021-00770-6}
  {\bibfield  {journal} {\bibinfo  {journal} {Nat. Photonics}\ }\textbf
  {\bibinfo {volume} {15}},\ \bibinfo {pages} {272} (\bibinfo {year}
  {2021})}\BibitemShut {NoStop}%
\bibitem [{\citenamefont {Yanay}\ and\ \citenamefont
  {Clerk}(2018)}]{Yanay2018}%
  \BibitemOpen
  \bibfield  {author} {\bibinfo {author} {\bibfnamefont {Y.}~\bibnamefont
  {Yanay}}\ and\ \bibinfo {author} {\bibfnamefont {A.~A.}\ \bibnamefont
  {Clerk}},\ }\bibfield  {title} {\bibinfo {title} {Reservoir engineering of
  bosonic lattices using chiral symmetry and localized dissipation},\ }\href
  {https://doi.org/10.1103/PhysRevA.98.043615} {\bibfield  {journal} {\bibinfo
  {journal} {Phys. Rev. A}\ }\textbf {\bibinfo {volume} {98}},\ \bibinfo
  {pages} {043615} (\bibinfo {year} {2018})}\BibitemShut {NoStop}%
\bibitem [{\citenamefont {Keck}\ \emph {et~al.}(2018)\citenamefont {Keck},
  \citenamefont {Rossini},\ and\ \citenamefont {Fazio}}]{Keck2018}%
  \BibitemOpen
  \bibfield  {author} {\bibinfo {author} {\bibfnamefont {M.}~\bibnamefont
  {Keck}}, \bibinfo {author} {\bibfnamefont {D.}~\bibnamefont {Rossini}},\ and\
  \bibinfo {author} {\bibfnamefont {R.}~\bibnamefont {Fazio}},\ }\bibfield
  {title} {\bibinfo {title} {Persistent currents by reservoir engineering},\
  }\href {https://doi.org/10.1103/PhysRevA.98.053812} {\bibfield  {journal}
  {\bibinfo  {journal} {Phys. Rev. A}\ }\textbf {\bibinfo {volume} {98}},\
  \bibinfo {pages} {053812} (\bibinfo {year} {2018})}\BibitemShut {NoStop}%
\bibitem [{\citenamefont {Damanet}\ \emph {et~al.}(2019)\citenamefont
  {Damanet}, \citenamefont {Mascarenhas}, \citenamefont {Pekker},\ and\
  \citenamefont {Daley}}]{Damanet2019}%
  \BibitemOpen
  \bibfield  {author} {\bibinfo {author} {\bibfnamefont {F.}~\bibnamefont
  {Damanet}}, \bibinfo {author} {\bibfnamefont {E.}~\bibnamefont
  {Mascarenhas}}, \bibinfo {author} {\bibfnamefont {D.}~\bibnamefont
  {Pekker}},\ and\ \bibinfo {author} {\bibfnamefont {A.~J.}\ \bibnamefont
  {Daley}},\ }\bibfield  {title} {\bibinfo {title} {Controlling quantum
  transport via dissipation engineering},\ }\href
  {https://doi.org/10.1103/PhysRevLett.123.180402} {\bibfield  {journal}
  {\bibinfo  {journal} {Phys. Rev. Lett.}\ }\textbf {\bibinfo {volume} {123}},\
  \bibinfo {pages} {180402} (\bibinfo {year} {2019})}\BibitemShut {NoStop}%
\bibitem [{\citenamefont {Yanay}\ and\ \citenamefont
  {Clerk}(2020)}]{Yanay2020}%
  \BibitemOpen
  \bibfield  {author} {\bibinfo {author} {\bibfnamefont {Y.}~\bibnamefont
  {Yanay}}\ and\ \bibinfo {author} {\bibfnamefont {A.~A.}\ \bibnamefont
  {Clerk}},\ }\bibfield  {title} {\bibinfo {title} {Reservoir engineering with
  localized dissipation: Dynamics and prethermalization},\ }\href
  {https://doi.org/10.1103/PhysRevResearch.2.023177} {\bibfield  {journal}
  {\bibinfo  {journal} {Phys. Rev. Research}\ }\textbf {\bibinfo {volume}
  {2}},\ \bibinfo {pages} {023177} (\bibinfo {year} {2020})}\BibitemShut
  {NoStop}%
\bibitem [{\citenamefont {Hama}\ \emph
  {et~al.}(2018{\natexlab{a}})\citenamefont {Hama}, \citenamefont {Munro},\
  and\ \citenamefont {Nemoto}}]{hama2018relaxation}%
  \BibitemOpen
  \bibfield  {author} {\bibinfo {author} {\bibfnamefont {Y.}~\bibnamefont
  {Hama}}, \bibinfo {author} {\bibfnamefont {W.~J.}\ \bibnamefont {Munro}},\
  and\ \bibinfo {author} {\bibfnamefont {K.}~\bibnamefont {Nemoto}},\
  }\bibfield  {title} {\bibinfo {title} {Relaxation to negative temperatures in
  double domain systems},\ }\href
  {https://doi.org/10.1103/PhysRevLett.120.060403} {\bibfield  {journal}
  {\bibinfo  {journal} {Phys. Rev. Lett.}\ }\textbf {\bibinfo {volume} {120}},\
  \bibinfo {pages} {060403} (\bibinfo {year} {2018}{\natexlab{a}})}\BibitemShut
  {NoStop}%
\bibitem [{\citenamefont {Hama}\ \emph
  {et~al.}(2018{\natexlab{b}})\citenamefont {Hama}, \citenamefont {Yukawa},
  \citenamefont {Munro},\ and\ \citenamefont {Nemoto}}]{hama2018negative}%
  \BibitemOpen
  \bibfield  {author} {\bibinfo {author} {\bibfnamefont {Y.}~\bibnamefont
  {Hama}}, \bibinfo {author} {\bibfnamefont {E.}~\bibnamefont {Yukawa}},
  \bibinfo {author} {\bibfnamefont {W.~J.}\ \bibnamefont {Munro}},\ and\
  \bibinfo {author} {\bibfnamefont {K.}~\bibnamefont {Nemoto}},\ }\bibfield
  {title} {\bibinfo {title} {Negative-temperature-state relaxation and
  reservoir-assisted quantum entanglement in double-spin-domain systems},\
  }\href {https://doi.org/10.1103/PhysRevA.98.052133} {\bibfield  {journal}
  {\bibinfo  {journal} {Phys. Rev. A}\ }\textbf {\bibinfo {volume} {98}},\
  \bibinfo {pages} {052133} (\bibinfo {year} {2018}{\natexlab{b}})}\BibitemShut
  {NoStop}%
\bibitem [{\citenamefont {Fauzi}\ \emph {et~al.}(2021)\citenamefont {Fauzi},
  \citenamefont {Munro}, \citenamefont {Nemoto},\ and\ \citenamefont
  {Hirayama}}]{fauzi2021}%
  \BibitemOpen
  \bibfield  {author} {\bibinfo {author} {\bibfnamefont {M.}~\bibnamefont
  {Fauzi}}, \bibinfo {author} {\bibfnamefont {W.~J.}\ \bibnamefont {Munro}},
  \bibinfo {author} {\bibfnamefont {K.}~\bibnamefont {Nemoto}},\ and\ \bibinfo
  {author} {\bibfnamefont {Y.}~\bibnamefont {Hirayama}},\ }\bibfield  {title}
  {\bibinfo {title} {Double nuclear spin relaxation in hybrid quantum hall
  systems},\ }\href@noop {} {\bibfield  {journal} {\bibinfo  {journal}
  {arXiv:2101.04433}\ } (\bibinfo {year} {2021})}\BibitemShut {NoStop}%
\bibitem [{\citenamefont {Astner}\ \emph {et~al.}(2017)\citenamefont {Astner},
  \citenamefont {Nevlacsil}, \citenamefont {Peterschofsky}, \citenamefont
  {Angerer}, \citenamefont {Rotter}, \citenamefont {Putz}, \citenamefont
  {Schmiedmayer},\ and\ \citenamefont {Majer}}]{Astner2017}%
  \BibitemOpen
  \bibfield  {author} {\bibinfo {author} {\bibfnamefont {T.}~\bibnamefont
  {Astner}}, \bibinfo {author} {\bibfnamefont {S.}~\bibnamefont {Nevlacsil}},
  \bibinfo {author} {\bibfnamefont {N.}~\bibnamefont {Peterschofsky}}, \bibinfo
  {author} {\bibfnamefont {A.}~\bibnamefont {Angerer}}, \bibinfo {author}
  {\bibfnamefont {S.}~\bibnamefont {Rotter}}, \bibinfo {author} {\bibfnamefont
  {S.}~\bibnamefont {Putz}}, \bibinfo {author} {\bibfnamefont {J.}~\bibnamefont
  {Schmiedmayer}},\ and\ \bibinfo {author} {\bibfnamefont {J.}~\bibnamefont
  {Majer}},\ }\bibfield  {title} {\bibinfo {title} {Coherent coupling of remote
  spin ensembles via a cavity bus},\ }\href
  {https://doi.org/10.1103/PhysRevLett.118.140502} {\bibfield  {journal}
  {\bibinfo  {journal} {Phys. Rev. Lett.}\ }\textbf {\bibinfo {volume} {118}},\
  \bibinfo {pages} {140502} (\bibinfo {year} {2017})}\BibitemShut {NoStop}%
\bibitem [{\citenamefont {Norcia}\ \emph {et~al.}(2018)\citenamefont {Norcia},
  \citenamefont {Lewis-Swan}, \citenamefont {Cline}, \citenamefont {Zhu},
  \citenamefont {Rey},\ and\ \citenamefont {Thompson}}]{Norcia2018}%
  \BibitemOpen
  \bibfield  {author} {\bibinfo {author} {\bibfnamefont {M.~A.}\ \bibnamefont
  {Norcia}}, \bibinfo {author} {\bibfnamefont {R.~J.}\ \bibnamefont
  {Lewis-Swan}}, \bibinfo {author} {\bibfnamefont {J.~R.~K.}\ \bibnamefont
  {Cline}}, \bibinfo {author} {\bibfnamefont {B.}~\bibnamefont {Zhu}}, \bibinfo
  {author} {\bibfnamefont {A.~M.}\ \bibnamefont {Rey}},\ and\ \bibinfo {author}
  {\bibfnamefont {J.~K.}\ \bibnamefont {Thompson}},\ }\bibfield  {title}
  {\bibinfo {title} {Cavity-mediated collective spin-exchange interactions in a
  strontium superradiant laser},\ }\href
  {https://doi.org/10.1126/science.aar3102} {\bibfield  {journal} {\bibinfo
  {journal} {Science}\ }\textbf {\bibinfo {volume} {361}},\ \bibinfo {pages}
  {259} (\bibinfo {year} {2018})}\BibitemShut {NoStop}%
\bibitem [{\citenamefont {Minganti}\ \emph {et~al.}(2021)\citenamefont
  {Minganti}, \citenamefont {Macr\`{\i}}, \citenamefont {Settineri},
  \citenamefont {Savasta},\ and\ \citenamefont
  {Nori}}]{minganti2021dissipative}%
  \BibitemOpen
  \bibfield  {author} {\bibinfo {author} {\bibfnamefont {F.}~\bibnamefont
  {Minganti}}, \bibinfo {author} {\bibfnamefont {V.}~\bibnamefont
  {Macr\`{\i}}}, \bibinfo {author} {\bibfnamefont {A.}~\bibnamefont
  {Settineri}}, \bibinfo {author} {\bibfnamefont {S.}~\bibnamefont {Savasta}},\
  and\ \bibinfo {author} {\bibfnamefont {F.}~\bibnamefont {Nori}},\ }\bibfield
  {title} {\bibinfo {title} {Dissipative state transfer and maxwell's demon in
  single quantum trajectories: Excitation transfer between two noninteracting
  qubits via unbalanced dissipation rates},\ }\href
  {https://doi.org/10.1103/PhysRevA.103.052201} {\bibfield  {journal} {\bibinfo
   {journal} {Phys. Rev. A}\ }\textbf {\bibinfo {volume} {103}},\ \bibinfo
  {pages} {052201} (\bibinfo {year} {2021})}\BibitemShut {NoStop}%
\bibitem [{\citenamefont {Carmichael}(2013)}]{carmichael2013statistical}%
  \BibitemOpen
  \bibfield  {author} {\bibinfo {author} {\bibfnamefont {H.~J.}\ \bibnamefont
  {Carmichael}},\ }\href@noop {} {\emph {\bibinfo {title} {Statistical methods
  in quantum optics 1: master equations and Fokker-Planck equations}}}\
  (\bibinfo  {publisher} {Springer Science \& Business Media},\ \bibinfo {year}
  {2013})\BibitemShut {NoStop}%
\bibitem [{\citenamefont {Li}\ \emph {et~al.}(2021)\citenamefont {Li},
  \citenamefont {Soret},\ and\ \citenamefont {Ciuti}}]{Li2021}%
  \BibitemOpen
  \bibfield  {author} {\bibinfo {author} {\bibfnamefont {Z.}~\bibnamefont
  {Li}}, \bibinfo {author} {\bibfnamefont {A.}~\bibnamefont {Soret}},\ and\
  \bibinfo {author} {\bibfnamefont {C.}~\bibnamefont {Ciuti}},\ }\bibfield
  {title} {\bibinfo {title} {Dissipation-induced antiferromagneticlike
  frustration in coupled photonic resonators},\ }\href
  {https://doi.org/10.1103/PhysRevA.103.022616} {\bibfield  {journal} {\bibinfo
   {journal} {Phys. Rev. A}\ }\textbf {\bibinfo {volume} {103}},\ \bibinfo
  {pages} {022616} (\bibinfo {year} {2021})}\BibitemShut {NoStop}%
\bibitem [{\citenamefont {Lee}\ \emph {et~al.}(2014)\citenamefont {Lee},
  \citenamefont {Chan},\ and\ \citenamefont {Wang}}]{Lee2014}%
  \BibitemOpen
  \bibfield  {author} {\bibinfo {author} {\bibfnamefont {T.~E.}\ \bibnamefont
  {Lee}}, \bibinfo {author} {\bibfnamefont {C.-K.}\ \bibnamefont {Chan}},\ and\
  \bibinfo {author} {\bibfnamefont {S.}~\bibnamefont {Wang}},\ }\bibfield
  {title} {\bibinfo {title} {Entanglement tongue and quantum synchronization of
  disordered oscillators},\ }\href {https://doi.org/10.1103/PhysRevE.89.022913}
  {\bibfield  {journal} {\bibinfo  {journal} {Phys. Rev. E}\ }\textbf {\bibinfo
  {volume} {89}},\ \bibinfo {pages} {022913} (\bibinfo {year}
  {2014})}\BibitemShut {NoStop}%
\bibitem [{\citenamefont {Xu}\ \emph {et~al.}(2014)\citenamefont {Xu},
  \citenamefont {Tieri}, \citenamefont {Fine}, \citenamefont {Thompson},\ and\
  \citenamefont {Holland}}]{Xu2014}%
  \BibitemOpen
  \bibfield  {author} {\bibinfo {author} {\bibfnamefont {M.}~\bibnamefont
  {Xu}}, \bibinfo {author} {\bibfnamefont {D.~A.}\ \bibnamefont {Tieri}},
  \bibinfo {author} {\bibfnamefont {E.~C.}\ \bibnamefont {Fine}}, \bibinfo
  {author} {\bibfnamefont {J.~K.}\ \bibnamefont {Thompson}},\ and\ \bibinfo
  {author} {\bibfnamefont {M.~J.}\ \bibnamefont {Holland}},\ }\bibfield
  {title} {\bibinfo {title} {Synchronization of two ensembles of atoms},\
  }\href {https://doi.org/10.1103/PhysRevLett.113.154101} {\bibfield  {journal}
  {\bibinfo  {journal} {Phys. Rev. Lett.}\ }\textbf {\bibinfo {volume} {113}},\
  \bibinfo {pages} {154101} (\bibinfo {year} {2014})}\BibitemShut {NoStop}%
\bibitem [{\citenamefont {Quach}\ and\ \citenamefont
  {Munro}(2020)}]{quach2020using}%
  \BibitemOpen
  \bibfield  {author} {\bibinfo {author} {\bibfnamefont {J.~Q.}\ \bibnamefont
  {Quach}}\ and\ \bibinfo {author} {\bibfnamefont {W.~J.}\ \bibnamefont
  {Munro}},\ }\bibfield  {title} {\bibinfo {title} {Using dark states to charge
  and stabilize open quantum batteries},\ }\href
  {https://doi.org/10.1103/PhysRevApplied.14.024092} {\bibfield  {journal}
  {\bibinfo  {journal} {Phys. Rev. Applied}\ }\textbf {\bibinfo {volume}
  {14}},\ \bibinfo {pages} {024092} (\bibinfo {year} {2020})}\BibitemShut
  {NoStop}%
\end{thebibliography}

\begin{thebibliography}{1}%
\makeatletter
\providecommand \@ifxundefined [1]{%
 \@ifx{#1\undefined}
}%
\providecommand \@ifnum [1]{%
 \ifnum #1\expandafter \@firstoftwo
 \else \expandafter \@secondoftwo
 \fi
}%
\providecommand \@ifx [1]{%
 \ifx #1\expandafter \@firstoftwo
 \else \expandafter \@secondoftwo
 \fi
}%
\providecommand \natexlab [1]{#1}%
\providecommand \enquote  [1]{``#1''}%
\providecommand \bibnamefont  [1]{#1}%
\providecommand \bibfnamefont [1]{#1}%
\providecommand \citenamefont [1]{#1}%
\providecommand \href@noop [0]{\@secondoftwo}%
\providecommand \href [0]{\begingroup \@sanitize@url \@href}%
\providecommand \@href[1]{\@@startlink{#1}\@@href}%
\providecommand \@@href[1]{\endgroup#1\@@endlink}%
\providecommand \@sanitize@url [0]{\catcode `\\12\catcode `\$12\catcode
  `\&12\catcode `\#12\catcode `\^12\catcode `\_12\catcode `\%12\relax}%
\providecommand \@@startlink[1]{}%
\providecommand \@@endlink[0]{}%
\providecommand \url  [0]{\begingroup\@sanitize@url \@url }%
\providecommand \@url [1]{\endgroup\@href {#1}{\urlprefix }}%
\providecommand \urlprefix  [0]{URL }%
\providecommand \Eprint [0]{\href }%
\providecommand \doibase [0]{https://doi.org/}%
\providecommand \selectlanguage [0]{\@gobble}%
\providecommand \bibinfo  [0]{\@secondoftwo}%
\providecommand \bibfield  [0]{\@secondoftwo}%
\providecommand \translation [1]{[#1]}%
\providecommand \BibitemOpen [0]{}%
\providecommand \bibitemStop [0]{}%
\providecommand \bibitemNoStop [0]{.\EOS\space}%
\providecommand \EOS [0]{\spacefactor3000\relax}%
\providecommand \BibitemShut  [1]{\csname bibitem#1\endcsname}%
\let\auto@bib@innerbib\@empty
\bibitem [{\citenamefont {Hama}\ \emph
  {et~al.}(2018{\natexlab{b}})\citenamefont {Hama}, \citenamefont {Yukawa},
  \citenamefont {Munro},\ and\ \citenamefont {Nemoto}}]{hama2018negative2}%
  \BibitemOpen
  \bibfield  {author} {\bibinfo {author} {\bibfnamefont {Y.}~\bibnamefont
  {Hama}}, \bibinfo {author} {\bibfnamefont {E.}~\bibnamefont {Yukawa}},
  \bibinfo {author} {\bibfnamefont {W.~J.}\ \bibnamefont {Munro}},\ and\
  \bibinfo {author} {\bibfnamefont {K.}~\bibnamefont {Nemoto}},\ }\bibfield
  {title} {\bibinfo {title} {Negative-temperature-state relaxation and
  reservoir-assisted quantum entanglement in double-spin-domain systems},\
  }\href {https://doi.org/10.1103/PhysRevA.98.052133} {\bibfield  {journal}
  {\bibinfo  {journal} {Phys. Rev. A}\ }\textbf {\bibinfo {volume} {98}},\
  \bibinfo {pages} {052133} (\bibinfo {year} {2018}{\natexlab{b}})}\BibitemShut
  {NoStop}%
\end{thebibliography}
\end{document}